\documentclass[12pt,preprint]{aastex}

\shorttitle{Optical Images and Source Catalog of {\it AKARI} NEP-Wide Survey Field}
\shortauthors{Jeon et al.}

\begin{document}

\title{Optical Images and Source Catalog \\
	of {\it AKARI} North Ecliptic Pole Wide Survey Field}

\author{Yiseul Jeon\altaffilmark{1,2}, Myungshin Im\altaffilmark{1,2}, 
Mansur Ibrahimov\altaffilmark{3},
Hyung Mok Lee\altaffilmark{2}, 
Induk Lee\altaffilmark{1,2,4}, 
and 
Myung Gyoon Lee\altaffilmark{2} }

\email{ysjeon@astro.snu.ac.kr \& mim@astro.snu.ac.kr}

\altaffiltext{1}{Center for the Exploration of the Origin of the Universe (CEOU), 
 Astronomy Program,  Department of Physics \& Astronomy,  Seoul National University,  
 Shillim-Dong,  Kwanak-Gu,  Seoul 151-742,  Republic of Korea
}
\altaffiltext{2}{Astronomy Program, FPRD, Department of Physics \& Astronomy,  Seoul National University,  
 Shillim-Dong,  Kwanak-Gu,  Seoul 151-742,  Republic of  Korea
}
\altaffiltext{3}{Ulugh Beg Astronomical Institute,  33 Astronomicheskaya str.,  Tashkent,  100052,  Uzbekistan
}
\altaffiltext{4}{Graduate Institute of Astronomy, National Central University,
No.300 Jhongda Rd., Jhongli City, Taoyuan County 320, Taiwan
}

\begin{abstract}
We present the source catalog and the properties of the $B-, R-$, and $I-$band images
obtained to support the {\it AKARI}  North Ecliptic Pole Wide (NEP-Wide) survey. 
The NEP-Wide is an {\it AKARI} infrared imaging survey of the north ecliptic pole
covering a 5.8 deg$^2$ area over 2.5 -- 6 $\micron$ wavelengths. 
The optical imaging data were obtained 
at the Maidanak Observatory in Uzbekistan using 
the Seoul National University 4k $\times$ 4k Camera on the 1.5m telescope. 
These images cover 4.9 deg$^2$ where no deep optical imaging data are  available.
Our $B-,  R-$, and $I-$band data reach the depths of $\sim$23.4,  $\sim$23.1,  
and $\sim$22.3 mag (AB) at 5$\sigma$, respectively. 
The source catalog contains 96,460 objects in the $R-$band,  and 
the astrometric accuracy is about 0.15$\arcsec$ at 1$\sigma$ in each RA and Dec direction. 
These photometric data will be useful for many studies 
including identification of optical counterparts of the infrared sources 
detected by {\it AKARI},  analysis of their spectral energy distributions  
from optical through infrared,  
and the selection of interesting objects to understand the obscured galaxy evolution. 
\end{abstract}


\section{ INTRODUCTION }

  Recently, the observational study of the cosmic star formation history has been 
 at the center of our efforts to understand the evolution and the formation of galaxies. 
  This is because the observational data play a key role in constraining physical processes
 governing the evolutionary history of galaxies on which outcomes of the theoretical predictions
 are strongly dependent.  Many studies have been carried out to derive star formation rates in distant
 galaxies  \citep[e.g.,][]{goto10,lebo09,lefl09,redd08,shim07,im05},
 however, there is a major obstacle in such efforts - the dust extinction of
 the light coming from star formation. 

  Since the discovery of ultra-luminous infrared galaxies (ULIRGs) and luminous
 infrared galaxies (LIRGs) by the InfraRed Astronomical Satellite (IRAS),
 it has been recognized that a significant portion of the light from the star formation is missing 
 in the optical/ultraviolet but emitted in infrared (IR).  Therefore, statistical studies of 
 infrared luminous galaxies to reveal the hidden cosmic star formation activities were major themes
 of the more recent infrared space telescope missions such as the Infrared Space Observatory (ISO), 
 the {\it Spitzer} Space Telescope, and the {\it AKARI} Space Telescope.
  Such studies found that the star formation at higher redshifts
 ($z \sim 1$) are more obscured than at $z \sim 0$ \citep[e.g.,][]{lefl05,bell07,goto10}
  underlining the importance of  infrared-based studies of galaxies. 
  However, due to the lack of efficient coverage between 10 to 20 $\micron$ in the {\it Spitzer} imaging, 
 comparison of the same rest-frame mid-infrared (MIR) quantities over a wide redshift
 range is not possible to further examine the usefulness of the MIR star formation
 indicators at various redshifts and the correlation between the star formation
 and the relative strengths of MIR spectroscopic features made by molecules
 such as PolyAromatic Hydrocarbon (PAH) and silicates.
 The additional MIR coverage at 
 10 to 20 $\micron$ is also advantageous in studying the MIR-excess emission in early-type 
 galaxies \citep{ko09,shim10}, 
 and the identification of Active Galactic Nuclei \citep[AGN; e.g.,][]{lee07,huan09}.

The {\it AKARI} satellite is an infrared space telescope launched in February  2006. 
By offering a contiguous, wide-field imaging capability 
in the wavelength range between 2.5 and 26$\micron$,  
{\it AKARI} complements the capability of the {\it Spitzer} space telescope,  
and provides an unprecedented view of the universe, especially at 11 and 15$\micron$. 
Because of the Sun-synchronous orbit of {\it AKARI},  
deep observations can be done only at the North and the South ecliptic poles. 
Therefore,  the North Ecliptic Pole (NEP) survey is  
one of the important extragalactic surveys of {\it AKARI}. 
The NEP survey has two major programs,  NEP-Wide and NEP-Deep surveys. 
The NEP-Wide field covers 5.8 deg$^2$ centered at NEP, whereas 
the NEP-Deep field covers 0.38 deg$^2$ with deeper exposures than the NEP-Wide. 
Due to the wide field, the contiguous MIR wavelength coverage, and 
the low extinction value of $E(B-V)=0.047$ at the center of NEP \citep{schl98},
the NEP-Wide field is unique  for the studies of the extragalactic objects 
like infrared luminous galaxies and AGNs. 
For further information about NEP surveys,  
the readers can refer to \citet{mats06},  \citet{wada07}, and \citet{lee07, lee09}.

Optical imaging data can enhance the value of the unique {\it AKARI} NEP survey.
Since {\it AKARI} has a low spatial resolution 
(the full width at half maximum (FWHM) of the point spread function (PSF) is about 6$\arcsec$ at 15$\micron$), 
the optical imaging with a  higher resolution can help 
improving the astrometric accuracy and help deblending confused objects. 
Also, the optical bands are important to construct 
multi-wavelength spectral energy distributions (SEDs) of infrared sources such as galaxies and AGNs  to understand their properties. 
There already exist deep optical data covering 2 deg$^2$ of  the central region of the NEP-Wide field,  
obtained by the Canada-France-Hawaii Telescope (CFHT)  with $u^*,  g',  r',  i'$, and $z'$ filters \citep{hwan07}.
However,  there is no deep optical data for the remaining NEP-Wide area. 
The Digitized Sky Survey images are available,  
but neither their depths nor spatial resolutions are sufficient 
(the depth is about 20 mag at $R$ and 
the seeing is about 3$\arcsec$ with pixel scale of 1.0$\arcsec$ at the NEP area)
for identifying optical counterparts of many infrared sources. 
Hence,  we observed this remaining NEP-Wide area in optical $B,  R$, and $I$ filters. 
 The $BRI$ filters were chosen because 
 we needed a multiple color set at least three colors to construct the optical part of the SED of galaxy;
 one blue part, one red part, and a middle part between red and infrared bands. 
 Also, those Bessell filters are wider than the Sloan Digital Sky Survey (SDSS) filters, enabling 
 us to achieve a higher signal to noise ratio (S/N) per unit exposure time than the SDSS filters. 
% Therefore the Bessell $BRI$ filters are suitable for our large area optical survey. 

 Throughout the paper, we use a cosmology with $\Omega_M=0.3,\Omega_\Lambda=0.7,$ 
 and $H_0$ = 70 km s$^{-1}$Mpc$^{-1}$ \citep[e.g.,][]{im97}. 
 We also use the AB magnitude system.

\section{OBSERVATION}

\begin{figure}
\plotone{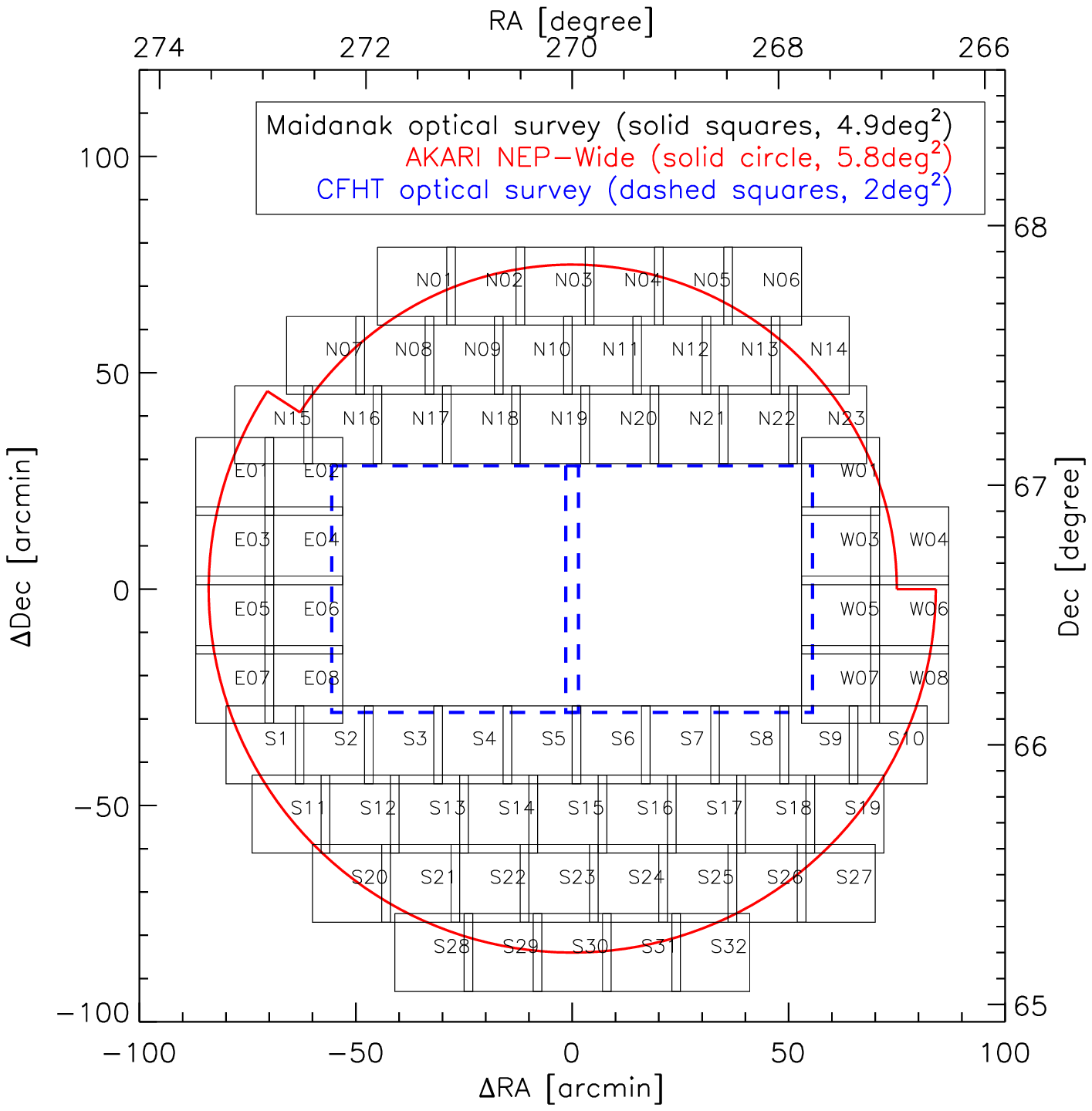}
\caption{Field coverage of the NEP-Wide optical survey. 
The center of the image corresponds to the NEP at $\alpha = 18^h00^m00^s,  \delta = +66\degr36\arcmin00\arcsec $.
Our $BRI$ imaging data (4.9 deg$^2$) are indicated with small squares,  while the thick, solid circle and 
the dashed squares indicate {\it AKARI} (5.8 deg$^2$) and CFHT (2 deg$^2$) coverage, respectively.
The $BRI$ images consist of 70 fields.
The names of the subfields and their central positions are listed in Table \ref{tbl0}.
\label{fig1}}
\end{figure}

The {\it AKARI} NEP-Wide field is centered at $\alpha = 18^h00^m00^s,  \delta = +66\degr36\arcmin00\arcsec $
covering a circular area of 5.8 deg$^2$. 
Our optical $BRI$ imaging survey covers 4.9 deg$^2$ 
outside the central 2 deg$^2$ area of the {\it AKARI} NEP-Wide.
Figure \ref{fig1} shows the field coverage of different components of the NEP-Wide survey. 
The thick, solid circle indicates the {\it AKARI} NEP-Wide (5.8 deg$^2$) coverage. 
The dashed squares are the CFHT optical survey area. 
The small squares show the Maidanak optical imaging area. 
It consists of 70 subfields,  each of them having an $18\arcmin\times18\arcmin$ field of view.  
These Maidanak fields are grouped into 4 different sections,  
East (8 fields),  West (7 fields),  North (23 fields),  and South (32 fields),  
and we name each subfield with E01--E08,  W01--W08,  N01--N23,  and S01--S32. 
The names of the subfields and their central positions are listed in Table \ref{tbl0}. 

The images were obtained in Bessell $B$, $R$, and $I$ filters 
from 2007 June 12 to August 5 at the Maidanak Observatory in Uzbekistan,  
using the Seoul National University 4k $\times$ 4k Camera (SNUCAM) on the 1.5 m Richey-Chretien, AZT-22 telescope \citep{im10}.  
The CCD camera has a 4096 $\times$ 4096 format and a pixel scale of 0.266$\arcsec$ at f/7.75. 
We used the four amplifier mode for the CCD image readout.  

Twilight flats were taken  at the start and the end of each night  when possible.
Since the $I-$band images are found to have complicated fringe patterns,
all of the $I-$band images were dithered by $\sim$20$\arcsec$ in random directions.
These dithered images were used to construct the master fringe image. 
For photometric calibration, we observed photometric standard fields 
from \citet{land92} at two different altitudes at least. 
In most subfields,  standard star images were obtained on the same night,  but in a few cases  
no corresponding standard star observations were carried out due to bad weather. 
In such cases,  the photometric calibration was done 
using a different way, as described in section \ref{sec:calib}.

\begin{figure}
\plotone{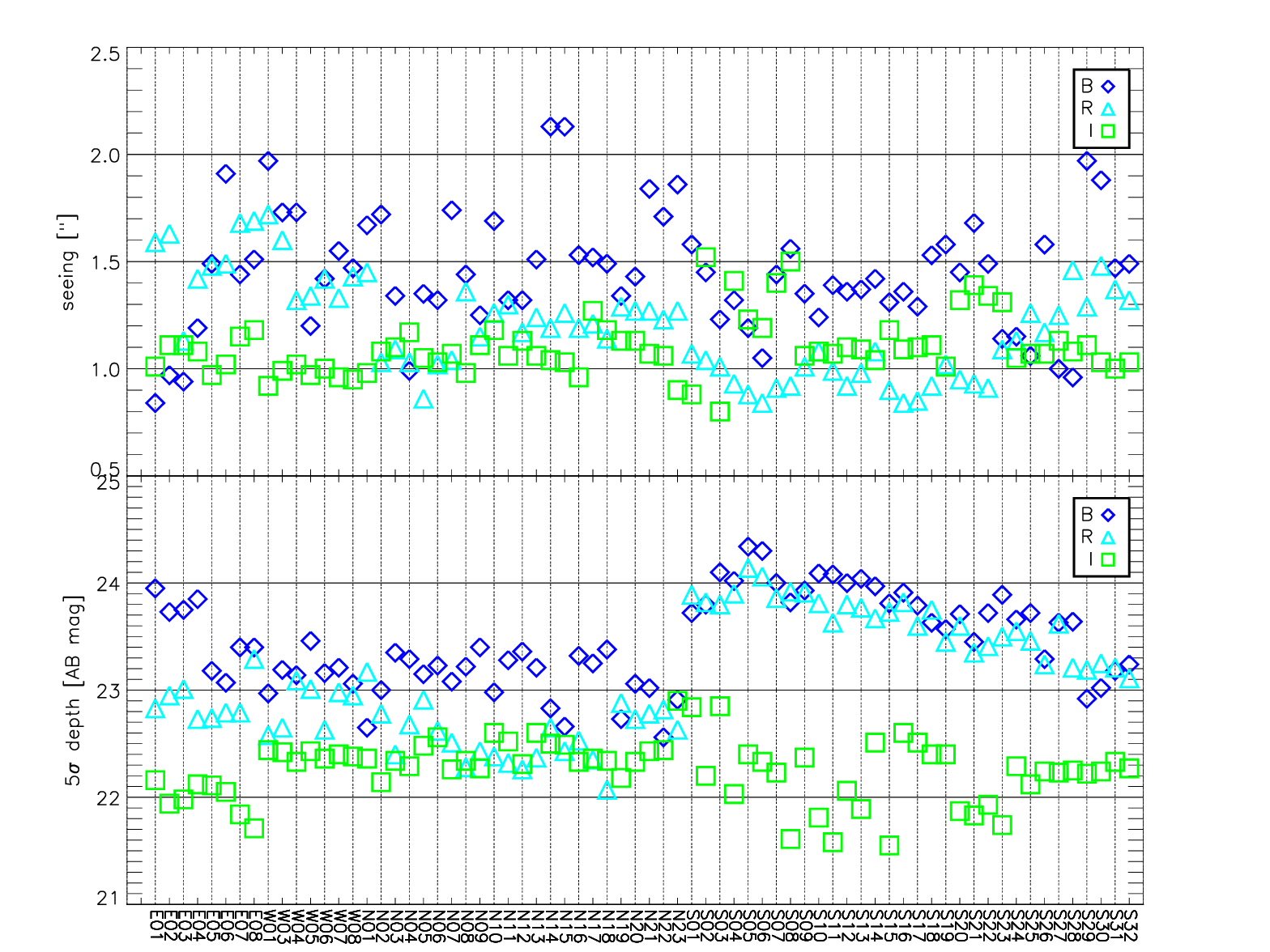}
\caption{
The upper and the lower panels show the seeing FWHMs and 
the 5$\sigma$ detection limits for each subfield, respectively
(the diamonds for $B$, the triangles for $R$, and the squares for $I-$band).
\label{fig2}}
\end{figure}

Table \ref{tbl1} summarizes our observations. 
For each filter,  typical total exposure times are 
21 min in $B$,  30 min in $R$,  and 20 min in $I$ filter. 
The images for a given subfield and a given filter were obtained mostly on a single night 
so that images are in homogeneous quality at least for a single filter of a given subfield. 
There are a few exceptions to this rule  
(3 subfields for $B-$band and 13 subfields for $I-$band).  
This happened when we needed to take images of a given subfield and a given filter over two nights 
due to observational constraints such as the lack of the observing time 
(e.g., the night was ending before finishing all the images of the subfield), 
or a rapidly deteriorating weather or seeing. 
 In some cases, the total exposure times and the number of images of  each subfield were adjusted to the
weather conditions. For example, when there were thin clouds in the sky, 
the exposure time or the number of images were increased to compensate for the loss of the S/N. 
The frame times for individual exposures were also adjusted. 
Shorter exposure times (2 minutes) were used when there were no guide stars, 
or the weather condition was varying too rapidly.
The $B-$ and $R-$band images were mostly taken during the same nights 
under dark/grey sky to provide images with similar seeing values, 
but  the $I-$band images were taken mostly during the bright time 
since the depth of $I-$band images is the least influenced by the presence of the moon among the $BRI$ filters.
The seeing condition for each subfield was
determined as the median value of FWHM of non-saturated point sources with the magnitudes between
13 and 15 mag.
In general, the seeing FWHM values of the stacked $R-$band and $I-$band images are stable 
at $\sim1.2\arcsec$ and  $\sim1.1\arcsec$, respectively, 
but there were more variations in the seeing conditions in $B-$band 
and the $B-$band images have the worst seeing FWHM values in general among the $BRI$ filters. 
 The maximum and the minimum seeing
 values among the $BRI$ filter images of a given subfield differ by 
 a factor of 1.5 in median, but in extreme cases, the difference
 could be as large as
 a factor of $\sim$2 due to deteriorated seeing conditions
 in the $B-$band.
The 5$\sigma$ detection limits of each subfield were determined from the sky background sigma,
within an aperture diameter of 3 times FWHM value. 
The detection limits vary from subfield to subfield, 
but the median 5$\sigma$ detection limit in $R-$band is 23.1 mag. 
Table \ref{tbl2} lists the seeings and 5$\sigma$ depths for each subfield. 
Figure \ref{fig2} shows the seeing and the 5$\sigma$ detection limit of the three filters for each subfield.

\section{DATA REDUCTION}
\subsection{Preprocessing}
We used IRAF\footnote{IRAF is distributed by the National Optical Astronomy Observatory,  
which is operated by the Association of Universities for Research in Astronomy,  Inc.,  
under cooperative agreement with the National Science Foundation.}   
\texttt{noao.imred.ccdred} package to process raw images. 
The data were processed with usual reduction procedures 
such as bias subtractions using a combined bias image 
and flat-field correction using twilight flat of each night. 
Dark correction was unnecessary since there is virtually no dark current 
in the images taken with our CCD camera. 

Because the bias subtraction and the flat fielding could not remove the amplifier bias level variations perfectly  
due to small temporal variations in the bias level,  
some of the preprocessed images show slightly different bias levels  in different quadrants, 
each representing a section of image read by a single amplifier.	 
To remove this pattern, we determined the background values of the 4 quadrants,
and subtracted the differences between the background levels of the 4 quadrant images from each quadrant. 
Then the background levels are normalized to the background level of a quadrant with the lowest background value.

We removed the fringe pattern in $I-$band following the steps below. 
First, we created a master fringe image combining all $I-$band images  (716 images) in median. 
Then, before subtracting the master fringe frame,  
we re-scaled the master fringe frame by an appropriate factor. 
This scaling factor is taken as the ratio of the median background value 
of each science image to the master fringe frame background value. 
Finally,  we subtracted the re-scaled master fringe frame from each $I-$band image. 
 We checked the difference in image quality before 
 and after the fringe removal.
  Before the fringe pattern removal, the large scale fringe pattern
 varies $\pm3\%$ of the background level which is comparable to the sky
 sigma value per pixel. After the fringe 
 subtraction, the large scale variation is reduced to 
 $< 0.5\%$ of the background or more than 6 times less
 than the sky sigma value, showing that the fringe pattern is
 removed successfully well below the sky sigma values.

After the fringe correction and the registration,  
all images were combined in median.
Cosmic rays are removed during the stacking of multiple images
with a sigma clipping adopting +3-sigma for rejecting outlier pixels.

\subsection{Astrometry}

  To derive astrometry solutions, we used a software named SCAMP 
 which finds astrometry solution by cross-matching automatically 
 an input reference catalog containing a set of accurate RA and Dec values 
 of sources against a list of object positions obtained from an input
 FITS image \citep{bert06}. 
  In our case, we used USNO-B1 catalog
 \citep{mone03} as the reference catalog for the astrometry calibration. 
  Astrometry solutions were derived 
 for all the stacked, subfield images in each filter separately. 
  The degree of the polynomial for distortion
 correction is adopted to be 4, since it gave the best solution among
 other options. SCAMP can refine the astrometry solution by 
 comparing positions of objects in overlapping images, but the procedure
 does not apply to our case.

\begin{figure}
\epsscale{1}
\plotone{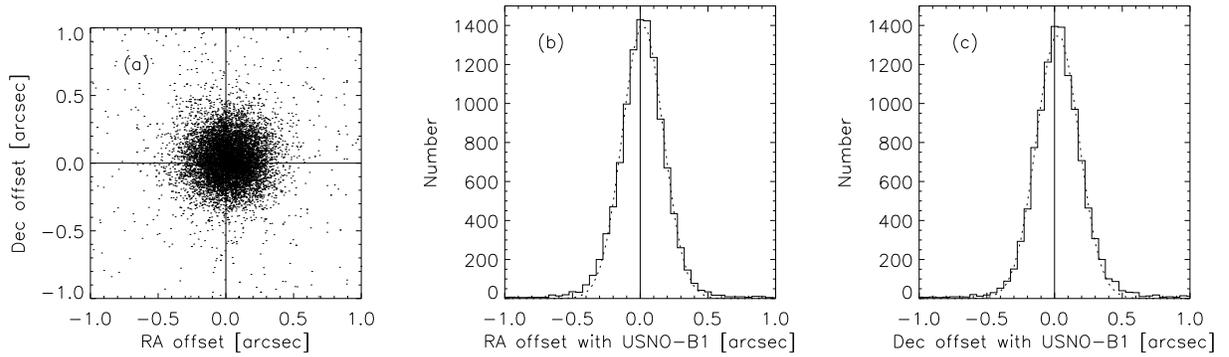}
\caption{Distribution (a) of the astrometry differences between our catalog and USNO-B1 catalog, 
and 	the histograms of the differences in RA (b) and  Dec (c) directions.
In (b) and (c), the dotted lines indicate the best-fit Gaussian functions 
to the  histograms of the distribution.
From the Gaussian function, 
the 1$\sigma$ astrometric errors are 0.16$\arcsec$ for RA and 0.15$\arcsec$ for Dec.
\label{fig3}}
\end{figure}

To check the accuracy of the astrometry solution, 
we compared our catalog with the USNO-B1 catalog stars 
that had the positional errors of $<$ 0.2$\arcsec$ and 
the apparent magnitudes of 14 $< B1$ mag $<$ 19. 
Figure \ref{fig3}-(a) shows the distribution of the differences in position
between the USNO-B1 stars and the corresponding objects in our catalog
in RA and Dec directions. 
Figures \ref{fig3}-(b) and (c) show the histograms of the astrometric differences 
between our catalog and USNO-B1 catalog of RA and Dec, respectively. 
To derive the astrometry error,
we fit the histogram of the astrometric differences with Gaussian functions. 
The dotted lines indicate the best-fit Gaussian function to the distributions.
After the fitting, 
we find that the 1$\sigma$ astrometric errors are 0.16$\arcsec$ for RA and 0.15$\arcsec$ for Dec.

\subsection{Photometric Calibration} \label{sec:calib}

We carried out the photometry calibration using three methods 
depending on the availability of the standard star data. 
The first method uses the Landolt standard stars \citep{land92}.
This procedure applies to  25 fields in $B-$band, 25 fields in $R-$band, and 18 fields in $I-$band,
where the standard star data were taken at two different altitudes at the least.
The adopted standard calibration formula is $m_{ins}-M_{std}=k_0+k_1X$,  
where $m_{ins}$ is an instrumental magnitude ( $m_{ins} = 25 - 2.5$ $\log$ $(DN/sec)$ ),  
$M_{std}$ is a standard magnitude,  $k_0$ is a zero point,  
$k_1$ is an airmass coefficient, and $X$ is an airmass. 
 The instrumental magnitudes were measured within an aperture of a 12$\arcsec$ diameter 
 on the standard stars. 
 We checked with growth curves that this aperture is large enough to contain virtually 
all photons 
from the star irrespective of the seeing of our data. 
We chose the standard stars that have similar colors for deriving the coefficients of the above formula.
To confirm the effect of not including the color terms, 
we checked the difference between the listed  magnitudes of standard stars 
with various colors (those not included in the derivation of the photometric zero-points)
 and the magnitudes of these standard stars derived with our photometry solutions.
 We find that the rms difference is less than 0.05 mag. That is, the magnitudes derived
 with our photometry solutions are independent of the color terms well within the error
 0.05 mag. This is consistent with the expected variation in the photometric zero points
 due to neglecting the color term, 
based on an observational campaign of photometry standards 
using the SNUCAM at Maidanak \citep{lim09}. 
Therefore we decided to neglect  the color term of the formula.
We found $k_0$ and $k_1$ values for each night and each filter,   
and derived the standard magnitudes of the objects using these values. 
The $k_0$ has the values in the range of 1.67 $\pm$ 0.13 at $B-$band, 
1.70 $\pm$ 0.11 at $R-$band, and 2.07 $\pm$ 0.08 at $I-$band.
Also, the $k_1$ have the values of 0.33 $\pm$ 0.05 at $B-$band, 
0.10 $\pm$ 0.03 at $R-$band, and 0.04 $\pm$ 0.01 at $I-$band.
These values are consistent with values derived from 
a campaign observation of standard stars \citep{lim09}.
 The zero-point errors were estimated from
the rms magnitude differences between the known magnitudes of the standard stars 
and the magnitudes of the same standard stars derived from our zero-point solution.
The zero-point error is about 0.032 mag using this method.

The second photometric calibration method is adopted for the data taken during the nights 
which have only one altitude data for standard stars
(34, 40, and 27 fields at $B-$, $R-$, and $I-$band, respectively.). 
In such cases, we used other night's $k_1$ value 
since the night-to-night variation in $k_1$ is not significant. 
This produces an additional zero-point error of 0.028 mag, at most. 
Consequently, the total error associated with the zero-point determination 
of the second photometric calibration method is less than 0.040 mag.

The third method of the photometric calibration is 
applied to the subfields without associated standard star observations.
For such fields (11, 5, and 25 fields in $B-$, $R-$, and $I-$band, respectively.), 
we used photometry of sources in the overlapping area of a neighbor subfield.
The corresponding sources that have magnitudes between 12 and 16 mag 
were used to measure the photometric zero-point.
For this method, the zero-point error was derived from
the standard deviation of the magnitude offset 
between the sources in the field where we want to determine the photometric zero-point 
and the matched sources in the neighbor fields.
In this case, the zero-point error is about 0.045 mag with respect to the photometry of the neighbor field. 
Therefore, the total error in the zero-point for the third method is about 0.032--0.045 mag, 
depending on the method used for the photometry calibration of the neighbor field.

We checked the consistency of our photometric calibration using an overlap area
between two neighboring fields
(those that were not used for the photometric calibration of the other field).
In the case of the first method, 
the mean photometry differences are $-0.006 \pm$ 0.028 mag in $B$ (30 fields for overlaps), 
0.002 $\pm$ 0.033 mag in $R$ (23 fields for overlaps), 
and 0.001 $\pm$ 0.017 mag in $I$ magnitudes (17 fields for overlaps).
In the case of the second method, the mean photometry differences are about 0.002 $\pm$ 0.030 mag in $B$ (37 fields for overlaps),
$-0.006 \pm$ 0.040 mag in $R$ (44 fields for overlaps), 
and $-0.022 \pm$ 0.035 mag in $I$ magnitudes (20 fields for overlaps).
Note that the consistency of our photometric calibration 
between different nights  are well below 0.05 mag for all filters
which is expected from the zero-point errors.

The $Z_p$ for each subfield are listed in Table \ref{tbl2}.
For the nights which have the $k_0$ and the $k_1$ values,  
the $Z_p$ means $25-(k_0+k_1X)$, 
i.e. the zero point for converting DN per exposure time to magnitude, 
including the airmass correction. 	
Therefore we can determine  the apparent  magnitude $M$, 
through $M=Z_p - 2.5\log$ $(DN/sec)$. 
The $Z_p$ can be used for the purpose of checking the depth of each subfield. 
Figure \ref{fig4} shows several depths of each subfield for the three filters. 
For each subfield, the depth are checked using the 5$\sigma$ detection limit,
the 50$\%$ completeness (see section \ref{sec:comple}), 
the 99$\%$ reliability (see section \ref{sec:reli}), 
or the $Z_p$, as well as the total exposure time.

\section{CATALOGS}

\subsection{Object Detection and Photometry} \label{sec:flags}

SExtractor \citep{bert96} was used for the detection and the photometry of objects in the images. 
We stacked $B-,  R-$, and $I-$band images together,  
and used the $BRI$ stacked image as our object detection image. 
 After many trials of SExtractor parameter sets and examining the detected
 and the non-detected sources by eye,
 we chose a set of 1.2$\sigma$ for DETECT\_THRESH,
 5 connected pixels for DETECT\_MINAREA, DEBLEND\_NTHRESH of 64,
 DEBLEND\_MINCONT of 0.005, 200 for BACK\_SIZE, and 3 for BACK\_FILTERSIZE 
 as the optimal set of detection parameters. 
 Lowering the detection limit produced too many spurious detections 
 and increasing the detection limit tend to miss obvious sources. 
 The adopted detection limit corresponds to the S/N of about 4.5 
 which is usually an optimal S/N limit for detection of objects
 \citep[e.g.,][]{shim06}.

 To cross-identify and perform the photometry of objects in each band,  
we used the ASSOC parameter in SExtractor with a matching radius of 0.658$\arcsec$, 
3$\sigma$ value of the astrometric error.
The above process may miss faint objects which are brighter in a particular filter
 than the others. 
To supplement the above source detection,  
we also ran SExtractor on images in each filter separately. 
This process reveals faint sources detected in a single filter image, 
but not detected in the $BRI$ stacked image. 
These sources detected in a single filter image were also matched 
with detections in other single filter images with the matching radius of 0.658$\arcsec$. 
Adding these faint sources,  we detected 
63,333 for $B$, 96,460 for $R$ and 70,492 sources for $I$ filter
over the reliability of 99$\%$ (see section \ref{sec:reli}). 
Among these objects, 16$\%$ in $B$, 33$\%$ in $R$, and 15$\%$ in $I$ are  
detected only in the single filter images. 
We derived aperture magnitudes using an aperture of diameter of 3 times FWHM 
and auto magnitudes with Kron-like elliptical apertures \citep{kron80}
which are taken as the total magnitudes . 
The aperture correction value is 0.10 $\pm$ 0.02 mag for
all the filters and all the subfields. 

 The magnitude errors are computed by combining the standard SExtractor estimates 
 and the zero-point errors we estimated.
 The standard SExtractor error estimates are based on the Poisson statistics, 
 and could underestimate the noise values when the sky noises in adjoining pixels are correlated  
 due to the stacking of the dithered frames \citep{gawi06}.
 Using a similar procedure described in \citet{gawi06},
 we calculated how the error within a square aperture of $N \times N$ pixels,  
 $\sigma_{N}$, changes as a function of $N$.
 In the case where each pixel noise is uncorrelated with each other, 
 $\sigma_{N} \propto N$. When the noise is correlated, the exponent of
 $N$ is greater than 1. In our data, we find 
 $\sigma_{N} \propto N^{1.14}$. For a magnitude measured
 within an aperture with 3.0$\arcsec$ diameter (3 times a typical FWHM of
 our image), there are about 110 pixels (or $N \simeq 10$). Therefore,
 the photometric errors of faint objects ($R \gtrsim 22$ mag) could be underestimated
 by as much as about a factor of 1.4. For the brighter objects
 ($R \lesssim 20$ mag), the photometric errors are not affected by
 this effect, since their errors are dominated by the object noise.

For the same object that exist on two or more fields in the overlapping areas,  
we chose an object that had the smallest FLAGS value from SExtractor 
(the FLAGS has a nonzero value if the object is blended, saturated,
 and/or truncated at an edge of a image)
and the smallest photometric error.
Among the same objects, we only included the one with a better S/N or image quality in the final catalog.
The examples of such cases are objects on the edge of an image or 
objects elongated due to astigmatism near the edge of the CCD in one image, 
but do not suffer such a problem in the other.
At the bright end,  the photometry is free from saturation and non-linearity above $R$ $\simeq$ 12 mag. 
We cut off the faint sources over the magnitude of false detection rate 1$\%$ 
(see section \ref{sec:reli}). 

We found spurious detections due to the reflection halos caused by the bright objects 
and due to faint, uncorrected fringe patterns in $R-$band images. 
In case of the reflection halos around bright stars, 
we divided the bright stars into three classes, 
and set a radius to define the region where spurious detections lie within it 
(120$\arcsec$ at 8 $\leq B$ mag $<$ 10 , 80$\arcsec$ at 10 $ \leq B$ mag $<$ 12 , 
40$\arcsec$ at 12 $ \leq B$ mag $<$ 14). 
We flag detections inside these radii as `near$\_$bright$\_$obj', 
since they are likely to be spurious detections or 
their photometry are not reliable.
We also found many spurious detections in the outer regions of stacked $R-$band images, 
thus we flag objects 
that are not matched with $B-$ or $I-$band catalogs and exist within 30$\arcsec$ from the edge
as `near$\_$edge'.
We found 5,186 (5$\%$) of `near$\_$bright$\_$obj' at $R-$band and 
14,833 (15$\%$)of `near$\_$edge'  sources, and checked those spurious sources by eyes.
These flags are included in the $BRI$ merged catalog.
Also we flag non-stellar sources as `galaxy' (see, section \ref{sec:galcut}). 
The flags are marked at the last column of the catalog (Table \ref{tbl3}).

\subsection{Star-Galaxy Separation} \label{sec:galcut}
To distinguish stellar sources from other extragalactic sources,  
we used a $B-H$ vs. $H-N2$ color-color diagram and the SExtractor stellarity index.
 Here, the stellarity indices measure the likelihood of a source 
 being a point source or an extended source, 
 with the value of 1 for a perfect point source, and 0 for a diffuse, extended source. 
 The stellarities are measured from the single-filter stacked images. 
 However, when we determine whether a source is  point-like or extended, 
 we adopted the stellarity value measured in a stacked image in the filter  
 where the photometric error of the source is the smallest. 
 Note that the stellarity values are nearly identical among different filter images 
 when an object is sufficiently bright ($R < 19$ mag). 
 We also note that the stellarity values are determined from the $R-$band images 
 for most of the objects, 
 since the $R-$band images generally have the highest S/N among the three filters.

The $H-$band  magnitudes come  from  the catalog made from the $J-$ and $H-$band 
images  of the NEP-Wide field  which were took with FLAMINGOS on the  2.1m telescope 
at Kitt Peak National Observatory (Jeon et al.,  in preparation),
and the $N2$ magnitudes come from the {\it AKARI} observation, 
of which the effective wavelength is 2.43$\micron$  \citep{lee07}.
We matched the objects detected in $H-$ and $N2-$band 
with the objects detected in $B-, R-$, and $I-$band 
using a matching radius of 1.5$\arcsec$ (for $H-$band) and 3$\arcsec$  (for $N2-$band), respectively. 

Figure \ref{fig5} shows  the $B-H$ vs. $H-N2$ diagram. 
\citet{bark_01} used $B-J$ vs. $J-K$ colors to distinguish quasars from stars. 
Instead of $J-K$ color, we used a similar selection method using $H-N2$.
On the left panel of Figure \ref{fig5},  the dots and the contours 
are made from the object detected in $B-$band with the magnitude error less than 0.1 
and matched with $H$ and $N2$  detections.
We can notice that there is a stellar locus at $H-N2$ $\simeq -0.5$ mag 
with 0 $<$ $B-H$ $<$ 5. 
Because most of stars have similar slopes at wavelength greater than 1$\micron$ on their SEDs, 
stars have $H-N2 \sim -0.5$ mag.
We checked the stellar locus using stars 
(the crosses of the right panel of Figure \ref{fig5}) in NEP-Wide field 
which had the stellarity greater than 0.8 and the magnitude of 10 $< H <$ 13.
Most of the stars are located at the stellar locus with a few exceptions. 

We also checked the location of galaxies using redshift tracks of galaxies.
The dashed line is for a star-forming galaxy, M51 \citep{silv98} and the dotted line is for a passively evolving galaxy
(see figure caption for the model used for this). 
We considered the intergalactic medium (IGM) attenuation \citep{mada96}  for all redshift tracks. 
Most of the galaxy track lie in the area of $H-N2$ greater than $-0.2$ mag, but
galaxies with z $\lesssim$ 0.1 lie at $H-N2 < -0.2$ mag. 
The PSF of galaxy at low redshift (z $<$ 0.1) is in general larger than that of star, 
so such sources can be identified as extended sources in the image. 
However, if the low redshift galaxies are faint and small, they may get misclassified as stars with the color cut.

Since the magnitude limit of the $H$ and $N2-$band are different from those in the optical, 
this method can be applied only to a relatively bright magnitude limit. 
Also, in order to identify bright, extended sources (namely, low redshift galaxies) 
in the stellar region on the color-color plot, 
we apply a stellarity cut where stellarity index has a value between 1 (point sources) and 0 (extended sources).

 In summary, we define stellar sources as the following: 
 \begin{center}
 Objects with $H-N2 < -0.2$ mag and the stellarity index $>$ 0.2 
 or \\
 Objects with $H-N2 \geq -0.2$ mag and the stellarity index $>$ 0.8  
 or \\
 if there is no $H-N2$ color available, stellarity index $>$ 0.8 
\end{center}
 
 The first condition weeds out extended objects at low redshift, 
 while the second condition allows us to select clear, point sources 
 in the region of galaxies on the color-color magnitude.
 The third criterion applies to a small number of objects near the detection limit of optical images,
 since $H$ and $N2$ images show detections of 98$\%$ and 77$\%$ of objects brighter than 
 the magnitude cut of the catalog at 50$\%$ completeness limit.

\subsection{Catalog Format}
The $B-, R-$, and $I-$band merged catalog is presented in Table \ref{tbl3}. 
The catalog contains objects that are deemed to be reliable above 99$\%$ confidence
where the confidence values are derived independently for each subfield in each filter
(see section \ref{sec:reli}).
Our catalog contains each object's identification number (ID), RA, Dec, 
the total magnitude and its error in each band,  
the aperture magnitude  within  a circular aperture with a diameter of 3 times FWHM  
and its error in each band,  the stellarity value from SExtractor,  
the galactic extinction value from the extinction map of \citet{schl98} for each band, 
and the flags described in section \ref{sec:flags}.
Note that the listed magnitudes are not corrected for the galactic extinction. 
The conversion formulas of Vega to AB magnitudes are $B($Vega$)=B($AB$)+0.09$,  
$R($Vega$)=R($AB$)-0.22$ and $I($Vega$)=I($AB$)-0.45$.
The offset values between Vega and AB magnitude 
are computed from the Vega spectrum \citep{bohl04}
and $B$, $R$, and $I$ filter response functions \citep{bess90} coupled with the CCD quantum efficiency curve.  
The magnitude error is the square root of the quadratic summation of the photometric measurement error 
and the zero-point error from the photometric calibration. 
We use a dummy value of 99.00 for non-detection.

Figure \ref{fig6} shows the stellarity distribution (the left three panels) 
and the photometric error distribution (the right three panels)
for all sources detected in $B-, R-$, and $I-$ band images.
The white lines on the right panels 
show the mode values for each magnitude bins, 
with the 25$\%$ and 75$\%$ quartiles indicated with error bars.
The spread of the data points in the magnitude versus error plots are 
caused mainly by varying depths of the data from different subfields. 
On average, the photometric error is less than 0.1 mag 
at $B$ $\lesssim$ 22.5 mag, $R$ $\lesssim$ 22 mag, and $I$ $\lesssim$ 21 mag.

\subsection{Completeness} \label{sec:comple}
Since the source detection near the detection limit is 
dominated by the detections performed on the single filter image 
($\simeq90\%$ of the faintest sources of $R-$band are detected only in the single filter image) 
rather than the $BRI$ stacked images, 
we computed the completeness using the $B-$, $R-$, and $I-$band 
single filter images of each subfield, 
to examine the completeness of the source detection. 
To estimate the completeness, we created
artificial objects for each magnitude bin from 12 to 23.5 mag 
and placed them randomly in the background regions of the single filter images.
The artificial objects are made of  100 point sources and 100 extended sources
where the elliptical galaxies account for 40$\%$ of the extended sources.
These artificial objects are created using the  \texttt{noao.artdata} package of the IRAF. 
After performing detection and photometry of these artificial sources 
with the same parameter setting as those applied to the actual images,
we calculated the completeness, 
which is defined as the ratio of the number of detected artificial sources 
to the number of input objects  for each magnitude bin.
Figure \ref{fig7} shows the completeness of $B-, R-$, and $I-$ band data. 
The gray solid lines in Figure \ref{fig7} represent the completeness of all the subfields
calculated with this method. 
We overploted the deepest (S05 in $B$, S05 in $R$ and N23 in $I$) 
and the shallowest (N22 in $B$, N18 in $R$, and S15 in $I$) fields in the 5$\sigma$ depths
with the black solid lines. 
The galaxy number counts (see section \ref{sec:galnum}) 
after the corrections based on these completeness result 
match the number counts from deeper surveys, meaning that our completeness estimates are reasonable.
Figure \ref{fig4} shows the 50$\%$ completeness as the asterisks for each subfield,
as well as the 5$\sigma$ depth, the 99$\%$ reliability (see section \ref{sec:reli}),
the $Z_p$, and the total exposure time. 
The magnitudes of completeness 50$\%$ for each subfield are presented  in Table  \ref{tbl2}.
We can see that the magnitudes at 50$\%$ completeness are brighter  than  
the ones of 5$\sigma$ detection limits  by a few tenths of magnitude.

\subsection{Reliability} \label{sec:reli}
We examined the reliability of our detections using false detection rates
for each filter and for each subfield. 
Here, we define false detection rate as the ratio of the number of objects 
detected in a negative image to the number of objects detected 
in the original image at a given magnitude bin. 
The negative image means the original image multiplied by $-1$. 
By this definition, an image with noises only would have the false detection rate of 1, 
while an image containing objects but with no noise would have the false detection rate of 0. 
When estimating the false detection rates, 
we excluded areas where objects are marked with `near\_edge',
since the inclusion of such area could disproportionately overestimate the false detection rate 
in the other areas which occupy most of the image.
 
\begin{figure}
\plotone{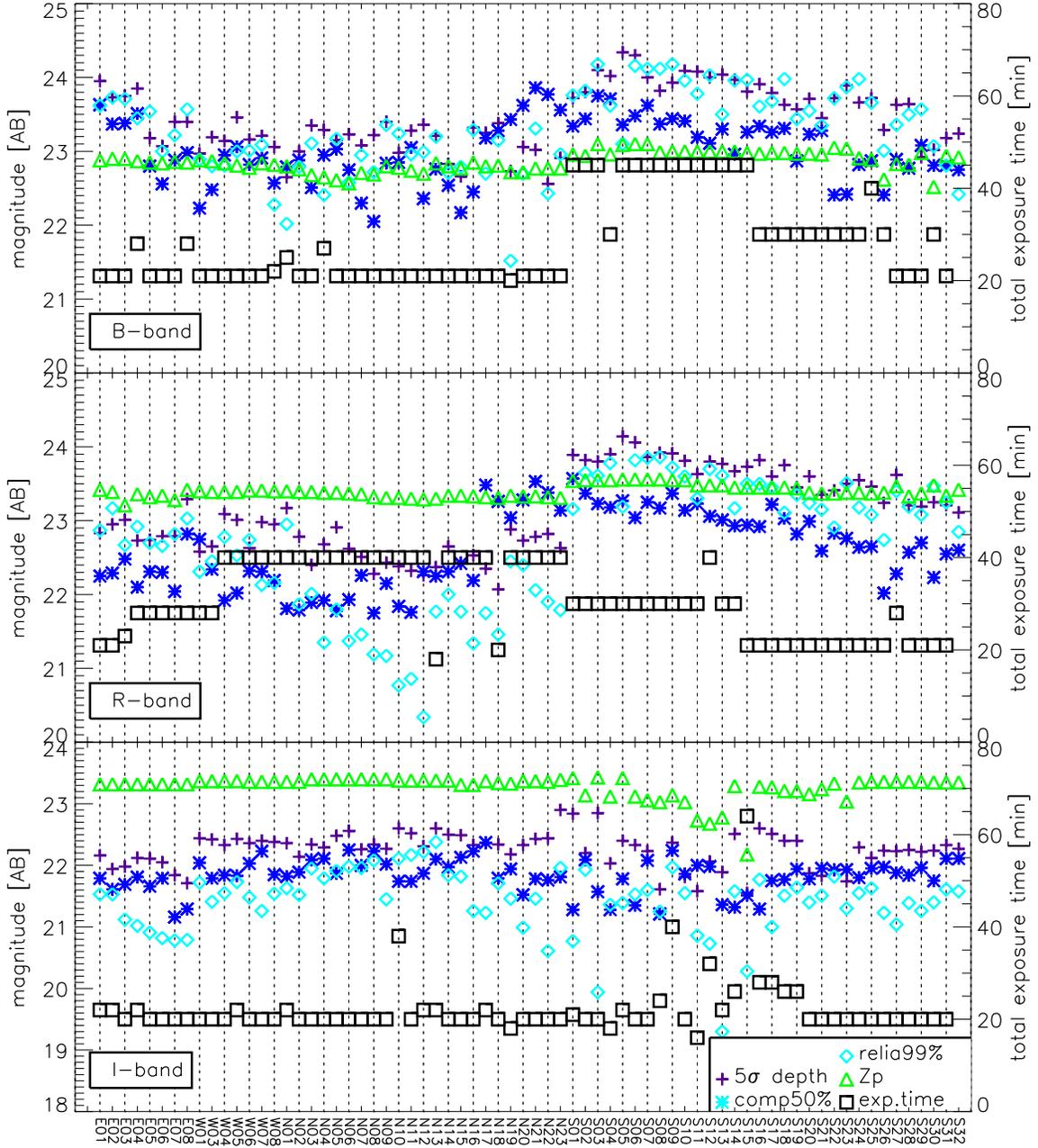}
\caption{
The figure shows several measures of the image quality and the depth of each subfield, 
such as the 5$\sigma$ depth (the cross symbols), 
the 50$\%$ completeness (the asterisks; see section \ref{sec:comple}), 
the 99$\%$ reliability (the diamonds; see section \ref{sec:reli}), 
the photometric zero point, $Z_p$ (the triangles), and the total exposure time (the squares). 
\label{fig4}}
\end{figure}

\begin{figure}
\plotone{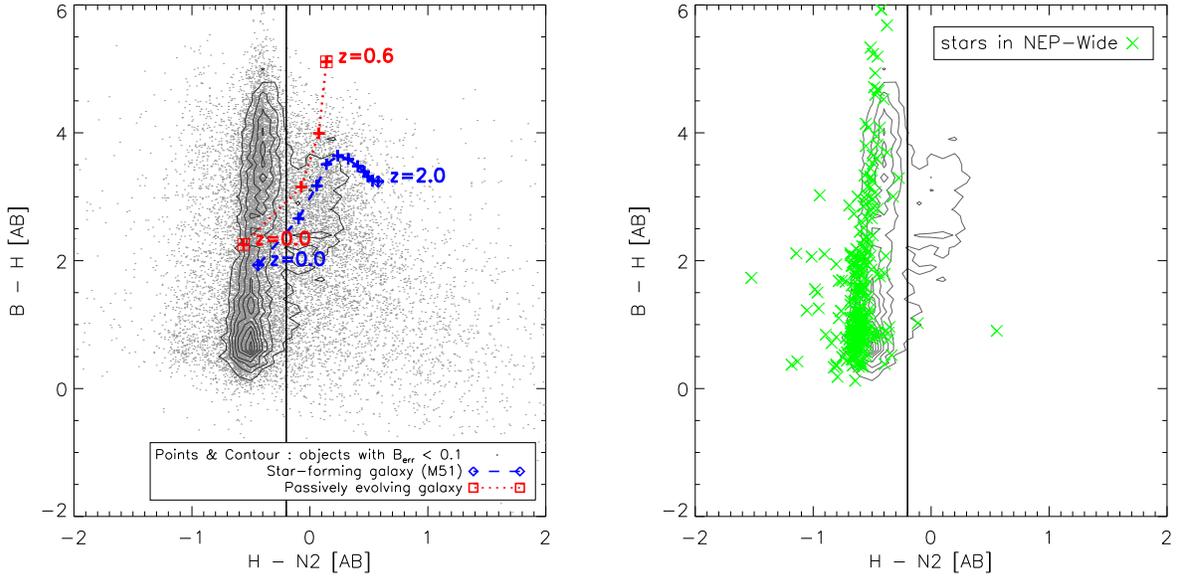}
\caption{ $B-H$ vs. $H-N2$ color-color diagrams and the density contours of sources detected in $B$, $H$, and $N2$.
The points and the contours in both figures are made with objects 
that have $B-$band magnitude errors less than 0.1 mag.  
The dashed line is a redshift track of a star-forming galaxy (M51) and 
the dotted line is for a passively evolving galaxy  
made with the \citet{bruz03} model 
of a passively evolving 5 Gyr-old galaxy with spontaneous burst, metallicity of Z=0.02, and  the Salpeter IMF. 
We use the vertical line as a boundary that distinguishes stellar sources against galaxies and AGNs.
In the right panel, we overlay stars in the NEP-Wide field (crosses)
to indicate the effectiveness of the color-cut in separating stars from galaxies.
\label{fig5}}
\end{figure}

\begin{figure}
\plotone{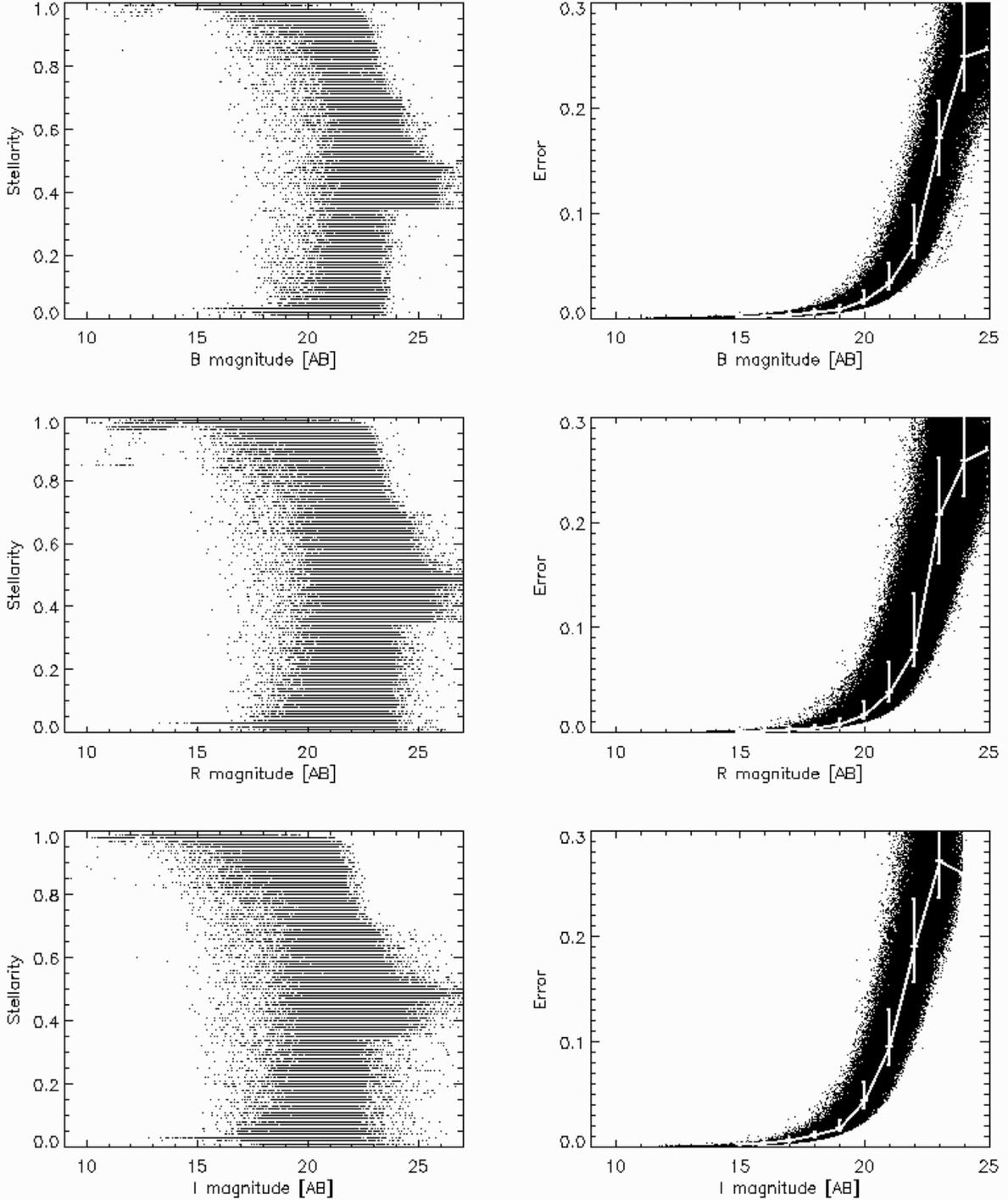}
\caption{SExtractor stellarity distribution and the photometric error distribution
for all sources detected in $B-, R-$, and $I-$ band images, respectively.
The stellarity index has a value between 1 (point sources) and 0 (extended sources).
In the right panel, we calculated the mode values of photometric error for each magnitude bin 
and let the 25$\%$ and 75$\%$ quartiles of the distributions as the error for each point. 
For $B$ $\lesssim$ 22.5 mag, $R$ $\lesssim$ 22 mag, and $I$ $\lesssim$ 21 mag, 
the photometric error is less than 0.1 mag.
\label{fig6}}
\end{figure}

\begin{figure}
\epsscale{.80}
\plotone{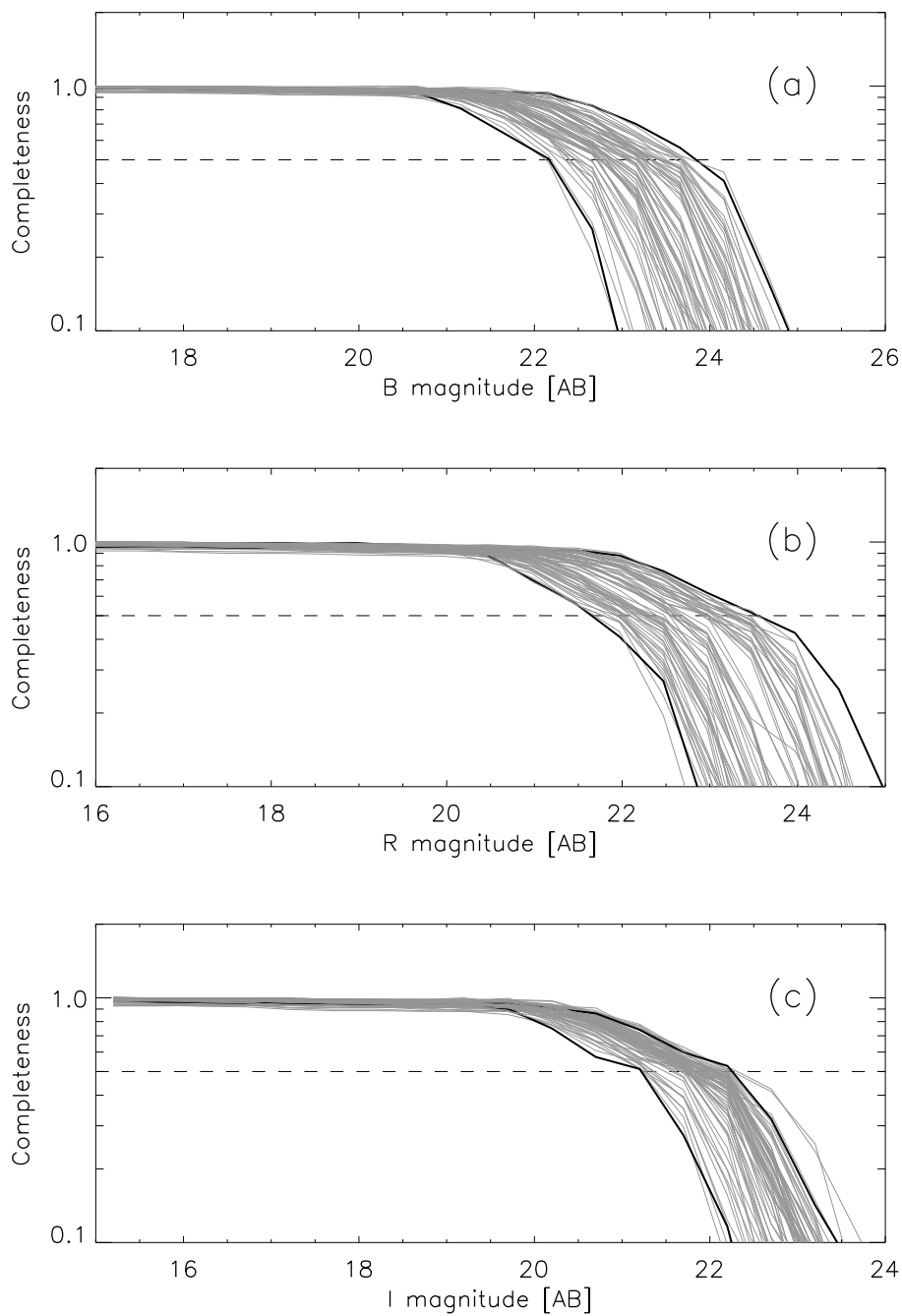}
\caption{Completeness for $B-, R-$, and $I-$band.
The gray solid lines are the completeness of all subfields from the simulation method, 
especially for the deepest fields 
(S05 for $B$, S05 for $R$, and N23 for $I$;the upper black solid lines) 
and the shallowest fields 
(N22 for $B$, N18 for $R$, and S15 for $I$;the lower black solid lines)
in the 5$\sigma$ depths. 
The horizontal dashed line is for 50$\%$ completeness cut. 
The magnitudes of completeness 50$\%$ for each subfield are shown in Table \ref{tbl2}.
\label{fig7}}
\end{figure}

\begin{figure}
\epsscale{.80}
\plotone{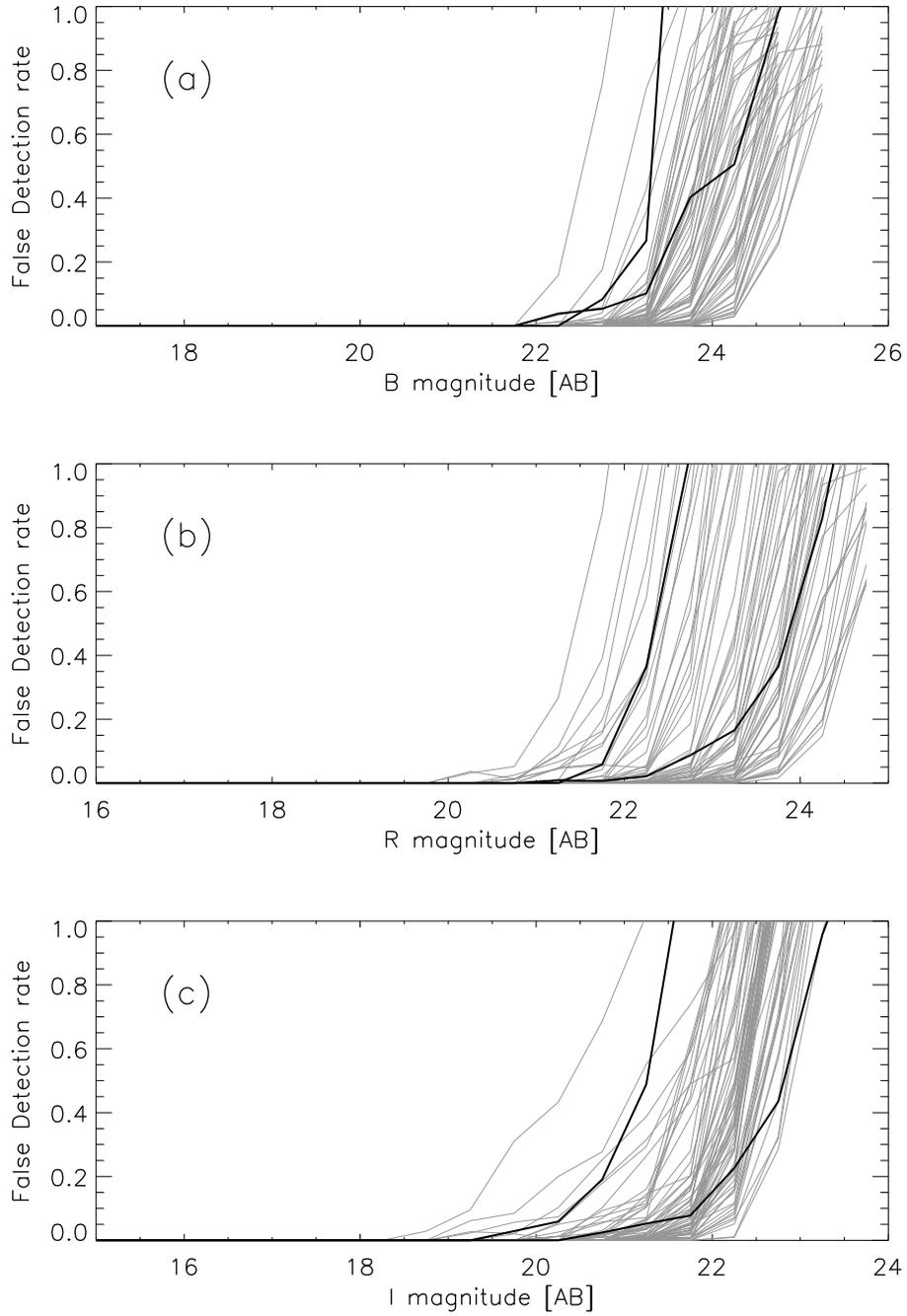}
\caption{False detection rate 
(the ratio of the number of objects detected in the negative image 
to the number of objects detected in the original image for each magnitude bin) of all subfields 
in $B-, R-$, and $I-$band. 
The black solid lines are for the deepest or the shallowest fields
as same as Figure \ref{fig7}. 
The magnitudes of false detection rate 1$\%$ for each field are shown in Table \ref{tbl2}.
\label{fig8}}
\end{figure}

Figure \ref{fig8} shows the false detection rates of $B-, R-$, and $I-$band data. 
The magnitude of the false detection rates 1$\%$ (or the reliability 99$\%$) 
of all the subfields are indicated in Table \ref{tbl2}.
Also, Figure \ref{fig4} shows the 99$\%$ reliability as diamond symbols for each subfield,
as well as the 5$\sigma$ depth, 50$\%$ completeness, the $Z_p$, and the total exposure time. 
Figure \ref{fig4} show that the magnitude limit of reliability 99$\%$
is greater than the 50 $\%$ completeness limit for $B-$ and $R-$band,
except $\simeq$20 fields.
 The 99$\%$ reliability limit in $I-$band is 
 consistently brighter than the 50$\%$ completeness limit by 0.3 mag. 
 The systematic trend arises from false detections of the fringe patterns left in $I-$band
image even after the fringe subtraction.
  Field-to-field variations of false detection rates and
 the completeness limits are much larger for $R$-band than 
 the other filters. 
  It is due to the large difference of the image depths of each subfield in $R-$band,
 where some fields (from N02 to N23) have relatively small depth with the 5$\sigma$ limit
 $<$ 22.5. It is because of the moon phase since these images were taken during full moon phase 
 (29 July -- 1 August). 
 
As a separate exercises, we examined the magnitude distribution of objects 
marked as `near\_edge' or `near$\_$bright$\_$obj'. 
Note that most of these objects are likely to be spurious detections. 
80$\%$ of these sources are found to be fainter than 
the magnitude limit of the 99$\%$ reliability. The 99$\%$ reliability
cut of the catalog removes most of the spurious detections,
but we note that objects marked
with `near\_edge' or `near$\_$bright$\_$obj' should be treated with caution.

\section{PROPERTIES OF THE DATA}
 
\subsection{Galaxy Number Counts} \label{sec:galnum}
To obtain the depth and homogeneity of this survey,  
we show the galaxy number count as a function of magnitude. 
The total area in which we used the galaxy number count of bright galaxies is 
about 3.8 deg$^2$ where $H-$ and $N2-$band data are also available. 
We take the Poisson error  as the error estimate of number counts. 

We calculated the number of galaxies per square degree 
at each magnitude bin. 
 Since the depth of each subfield is different from another,
 the combined number counts of all the NEP survey fields were 
 computed following the steps below.
 First, the number count was computed for each subfield up to
 the completeness limit of 50$\%$ as determined in section \ref{sec:comple}. 
 Then, the completeness correction was applied to the number count of each subfield. 
 Finally, we combined the number counts from all the subfields 
 to derive the final number count result. 
 With this method, the bright end of the number counts,
 usually susceptible to field-to-field variations, include the data from
 all the 70 fields, while the deepest number counts include the data from
 the deepest subfields only.
 We also calculated the number count of stars using the stellar objects of magnitude 
 brighter than the completeness limit of $\sim50\%$.

\begin{figure}
\epsscale{.80}
\plotone{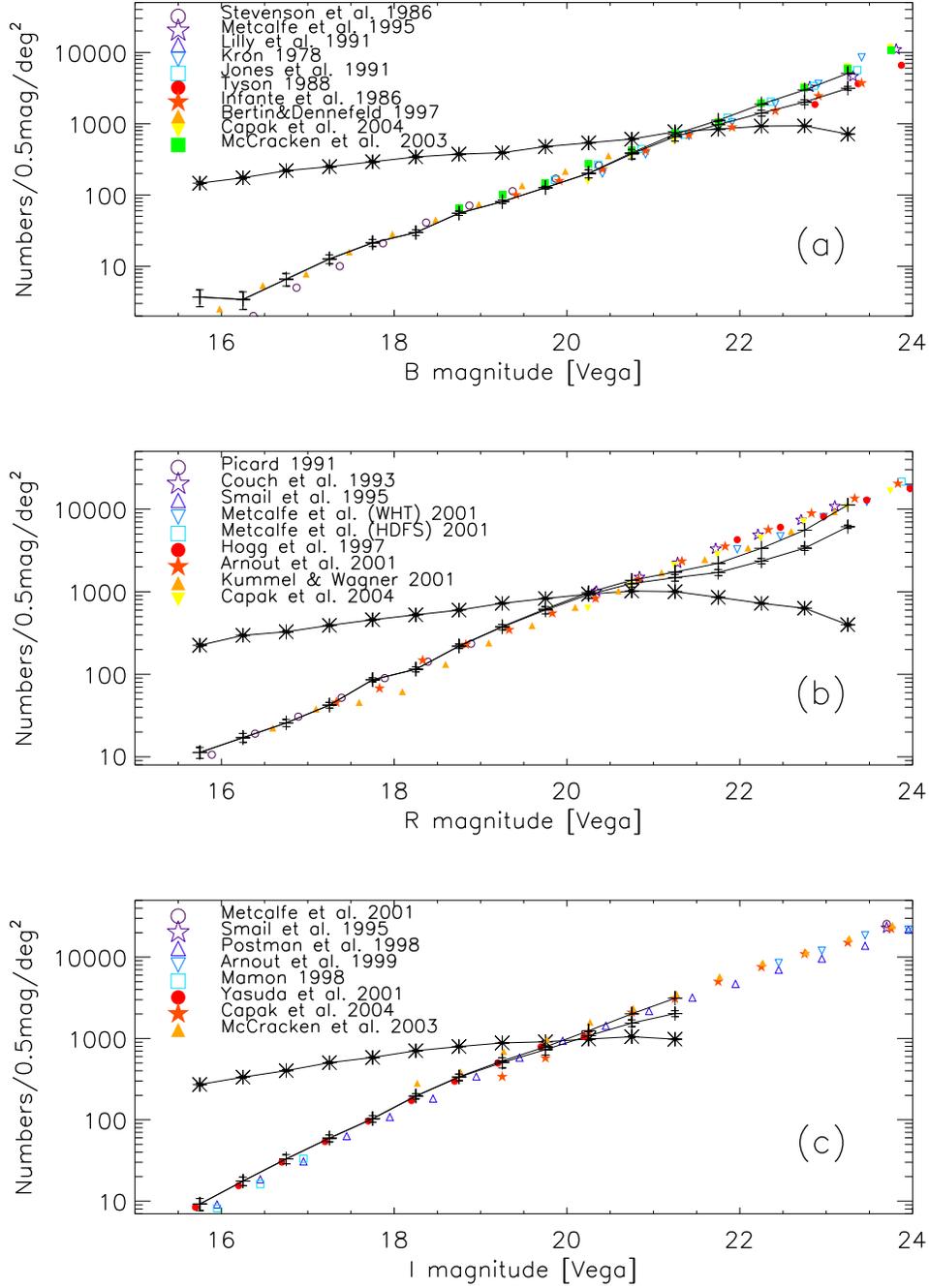}
\caption{Galaxy and stellar number counts of $B-, R-$, and $I-$band
(two solid lines with crosses).
The error estimates of number counts are the Poisson errors.
The lower solid line is for galaxy number count without completeness correction, and 
the upper solid line is from the galaxy number count corrected using the completeness. 
We compare our number counts with the number counts in the literature (other symbols).
The solid lines with asterisks are for the number count of stars.
The galaxy and stellar number count data are provided in Table \ref{tbl4}.
\label{fig9}}
\end{figure}

Figure \ref{fig9} displays the galaxy and the stellar number counts in $B-,  R-$, and $I-$band. 
The lower solid line with crosses represents raw number count and 
the upper solid line with crosses is the number count after applying correction for the completeness.
The solid line with asterisks shows the number count of stars.
Also we compare our number counts with those in the literature
\citep{arno99,arno01,bert97,capa04,couc93,hogg97,infa86,jone91,kron78,kumm01,lill91,
mamo98,mccr03,metc95,metc01,pica91,post98,smai95,stev86,tyso88,yasu01}.
Since some of the filter systems in the literature are slightly different from ours,  
we applied the magnitude transformation by \citet{metc95,metc01} to the literature values.
We can see that our number counts match with others.  
The galaxies and stars number count data are provided in Table \ref{tbl4}.

\subsection{ Color-Magnitude Diagram} \label{sec:cmd}
\begin{figure}
\epsscale{.80}
\plotone{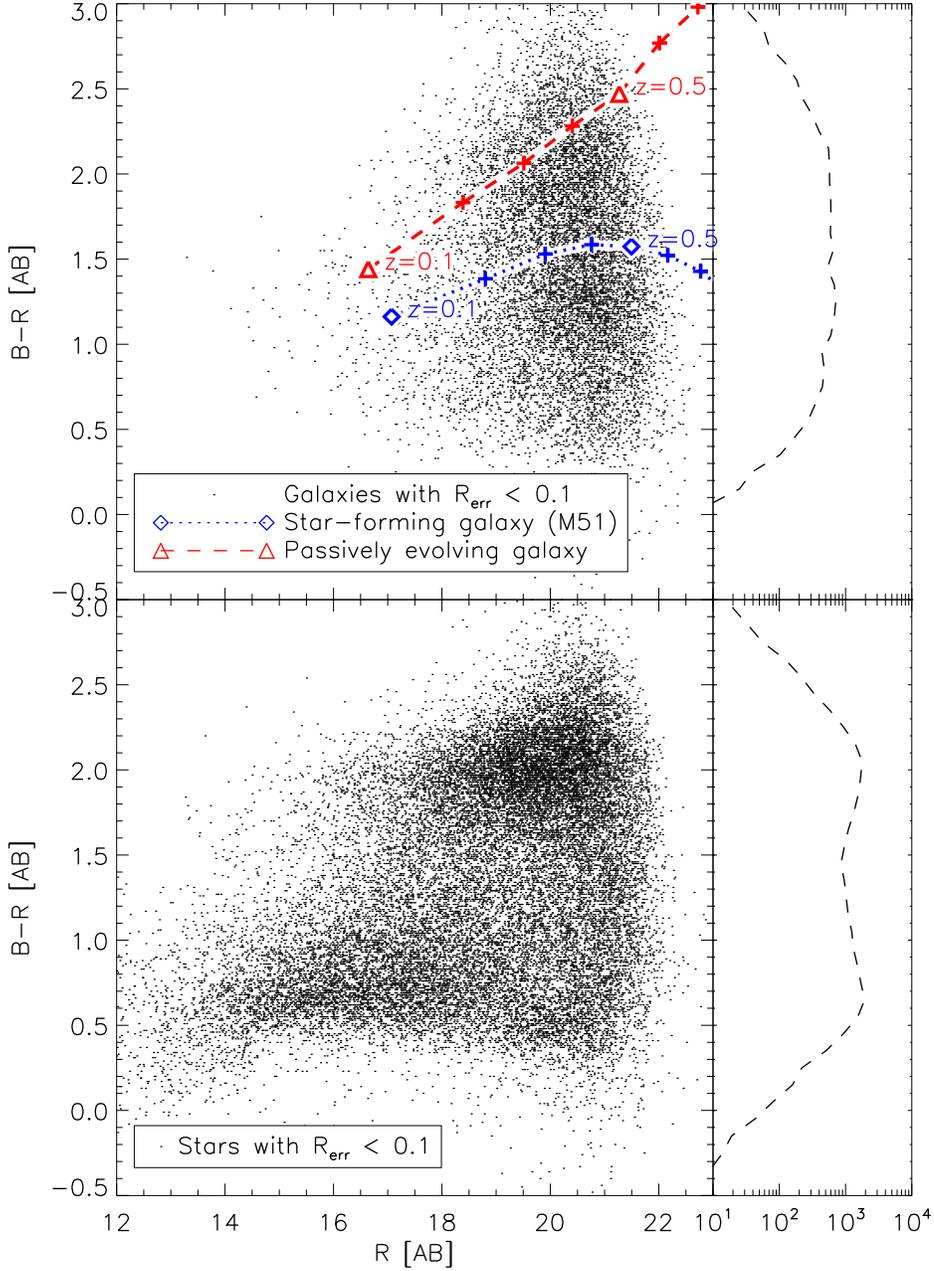}
\caption{The color-magnitude diagram of $R$ vs. $B-R$ (left) 
and histograms showing the distribution $B-R$ colors (right).
The upper panel shows the galaxies with $R$ magnitude error less than 0.1.
The dotted line is a redshift track  for a star-forming galaxy (M51) and the dashed line is for a passively evolving galaxy. 
The redshift track is for galaxies with the characteristic absolute magnitudes (see the text).
The lower panel shows the stars of the same conditions in the upper panel. 
\label{fig10}}
\end{figure}

\begin{figure}
\epsscale{.80}
\plotone{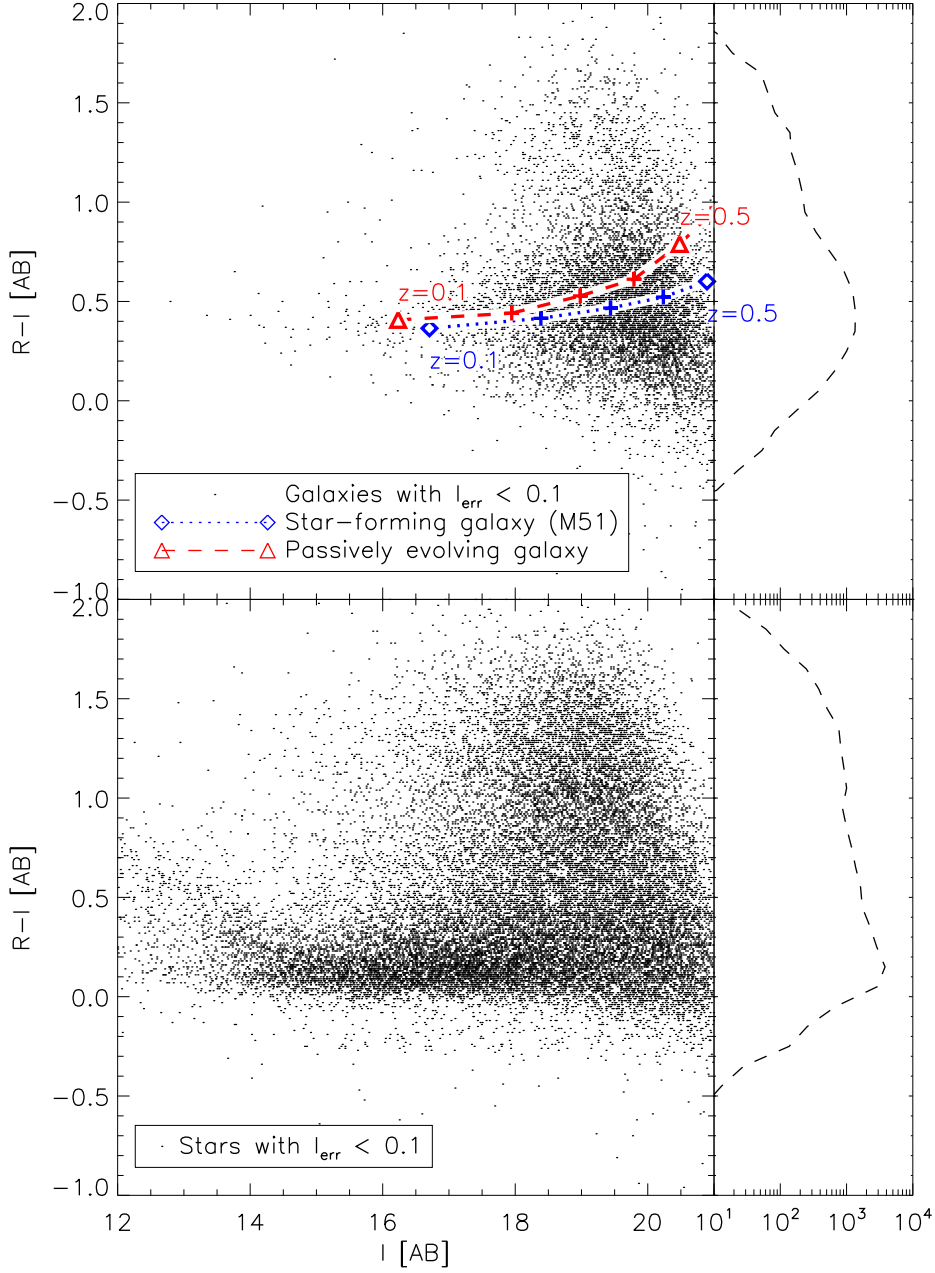}
\caption{ $I$ vs. $R-I$ color-magnitude diagrams (left) and the distribution of $R-I$ colors (right)
with the sources of $I$ magnitude error less than 0.1.
The upper panel is for the galaxies and the lower panel is for the stars. 
The symbols are the same as those in the Figure \ref{fig10}.
\label{fig11}}
\end{figure}

Figure \ref{fig10} displays the color-magnitude diagrams of $R$ vs. $B-R$  (left) and its histograms (right).
The upper panel shows the diagram for galaxies  
with $R$ magnitude error less than 0.1, and
the lower panel shows for the stars as the same conditions  with galaxies  in the upper panel. 
We checked positions of galaxies on the color-magnitude diagram traced by the redshifts.
The dotted line represents a redshift track of 
a star-forming galaxy, M51 \citep{silv98}, 
and the dashed line is for a passively evolving galaxy. 
The SED of a passively evolving galaxy is
for a 5 Gyr old stellar population, created with the \citet{bruz03} model, 
assuming  a spontaneous burst of single stellar population with 
metallicity of Z=0.02 and the Salpeter IMF.
We included the attenuation of galaxy light due to the IGM when computing the colors \citep{mada96}.
Two different lines represent galaxies with characteristic absolute magnitudes:
for an early type galaxy, 
the characteristic absolute magnitude, $M^*(r^*)$, is taken as $-20.75+5 \log h$, 
and for a late type galaxy, $-20.30+5 \log h$ is adopted for $M^*(r^*)$ \citep{naka03}.
To convert the $r^*-$band magnitude to the $R-$band magnitude,
we used the magnitude transformation relation of  \citet{jest05}, 
and $g-r \simeq 0.8 $ as the mean color of galaxies. 
The k corrections are calculated from the templates of two galaxies  in a standard manner. 
The $B-R$ colors of elliptical and spiral galaxies are 1--1.5 mag at z=0.1. 
For 0.2 $\leq$  z $\leq$ 0.6, the elliptical galaxies have the $B-R$ values of 1.5--3 mag and 
the spiral galaxies have the $B-R$ values of 1--1.5 mag.
The color-magnitude plot shows that our $B$ and $R-$band images 
detect galaxies at $z=0$ through $z=0.6$, 
and galaxies brighter than $M^*(r^*)$ to higher redshifts ($z \sim 1$). 
On the other hand, we find bimodality in the color-magnitude diagrams of stellar objects 
(bottom plot, Figure \ref{fig10}).
The bimodality originates from G dwarfs that dominate the color 
at $B-R$ $\simeq$ 0.5 and M Giant stars at $B-R$ $\simeq$ 2.
Figure \ref{fig11} displays the color-magnitude diagrams in $I$ vs. $R-I$ (left) and its histograms (right)
with the sources of the $I$ magnitude error less than 0.1.
The symbols and the tracks are the same as in Figure \ref{fig10}.
The $R-I$ color of elliptical and spiral galaxies is  0.3--0.5 at z $\lesssim$ 0.2 
and 0.5--0.9 at z $\simeq$ 0.5. 
Like Figure \ref{fig10}, the distribution of objects in the $I$ versus $R-I$ plot confirm that 
$I-$band images have the depth similar to the $R-$band images in terms of 
detecting galaxies at different redshifts.
We can notice that the model colors of $B-R$ and $R-I$ are 
consistent with the distributions of galaxies. 
We expect that the combination of the optical data and the {\it AKARI} infrared data 
to provide a unique view on the evolution of galaxies, 
such as the morphological transformation process through studying MID excess emission 
in early-type galaxies \citep{ko09}.

\section{Discussion}
Here, we investigate how our data compare with other similar optical datasets, 
and explore the potential of selecting high redshift galaxies and quasars using the data.

First, many extragalactic surveys have been performed with various depths and area coverage. 
The 5.8 deg$^2$ coverage of the NEP-Wide survey is wider than that of 
the Cosmic Evolution Survey \citep[COSMOS,][]{scov07}, 
or other deeper surveys such as DEEP2 \citep{davi03} 
and the Great Observatories Origins Deep Survey \citep[GOODS,][]{dick03}. 
On the other hand, the area coverage is smaller than 
that of the {\it Spitzer} Wide-Area Infrared Extragalactic Survey (SWIRE) survey \citep{Lons03} by a factor of 10 
with a similar depth in infrared as well as optical ancillary datasets.  
In terms of the depth and area coverage, our survey is 
most similar to the {\it Spitzer} First Look Survey \citep[FLS,][]{fadd04}. 
Our infrared depth is a bit shallower than those of the FLS, 
while we provide a contiguous wavelength coverage including 11 to 15 micron. 
Consequently, our optical imaging depths are designed to match the infrared survey depths, 
thus about a magnitude shallower than the $R-$band depth of the FLS ancillary data.

\begin{figure}
\epsscale{.90}
\plotone{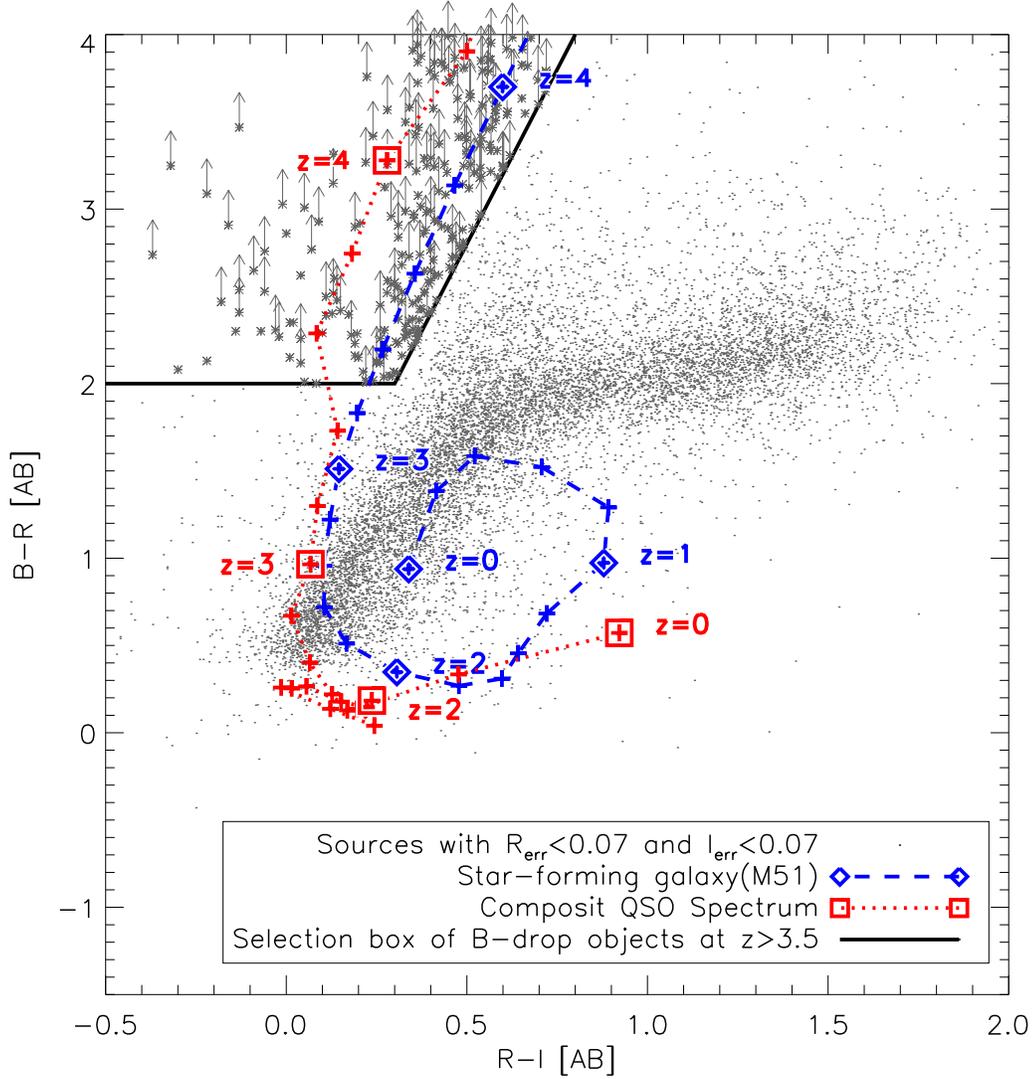}
\caption{The color-color diagrams of $B-R$ vs. $R-I$ and 
the redshift tracks of a star-forming galaxy and a quasar. 
The gray points are sources with $R$ and $I$ magnitude errors less than 0.07. 
The dashed line with diamond points represents a track for M51, and  
the dotted line with squares is for a redshift track
made from a median  composite spectrum of SDSS quasars from \citet{vand01}. 
The black boxed area in the upper-left corner defines the selection region 
for high redshift objects at z $>$ 3.5.
The asterisks show the sources in this area
and the asterisks with arrows indicate objects with 2$\sigma$ lower limits for $B-R$ colors, 
when the object was not detected in $B-$band.
\label{fig12}}
\end{figure}

Next, we discuss the usefulness of our datasets for selecting high redshift objects 
such as Lyman Break Galaxies (LBGs) and quasars.  
Figure \ref{fig12} shows color-color diagrams in $B-R$ vs. $R-I$. 
We plotted sources with $R$ and $I$ magnitude errors less than 0.07 mag as gray points
which correspond to objects with $R \lesssim$  21.0--23.0 mag, and $I \lesssim$ 20.4--21.8 mag. 
The dashed line with diamonds represents a redshift track for a star-forming galaxy \citep[M51,][]{silv98} and 
the dotted line with squares is a redshift track of a median composite spectrum 
of SDSS quasars from \citet{vand01}. 
The attenuation of the source light below Lyman $\alpha$ due to IGM 
is calculated using the model of \citet{mada96}.
Using the redshift tracks as a guide, 
we define a selection box in the color-color diagram 
where objects at z $>$ 3.5 can be found 
\citep[$B-$band dropouts; e.g.,][]{stei99,giav04,kang09,glik10}.
The selection box is defined with the following equations: 

\begin{mathletters}
\begin{eqnarray}
B-R & \geq & 2 \\
B-R & \geq & 4 (R-I) +0.8
\end{eqnarray}
\end{mathletters}

Although LBGs could be found in this selection box  in principle, 
the expected number of the $B-$dropout galaxies in our data is  very low. 
Extrapolating the surface number density of z $\sim$ 4 LBGs from \citet{stei99}
assuming an exponential decrease of the brightest objects as in the Schechter function, 
we expect about one or less than one $B-$dropout galaxies at $I <$ 21.5 mag. 
On the other hand, the number density  could be higher by a factor of 10 or so, 
if the bright end of the luminosity function flattens due to AGN contribution 
as seen in the luminosity functions of $U-$band dropout objects at $z \sim 3$ 
\citep[e.g.,][]{hunt04,shim07}.
In our data, we identify $\sim$800 $B-$dropout candidates in the selection box, 
but we believe that most of these are interlopers such as stars and low redshift galaxies
due to the large mismatch with the expected number.

The expected number of quasars over the 4.9 deg$^2$ is $\lesssim$ 10 at $I <$ 20 mag \citep{rich06},
 or 10 and 20 at $R <$ 21.5 mag and 22.5 mag, respectively \citep{glik10}.
  After inspecting images of the $B$-dropout candidates visually and examining 
 their multi-wavelength SEDs from the optical through the MIR 11$\micron$-band 
 of {\it AKARI}, we identify at least several objects that have the SED shapes consistent
 with those of high redshift quasars.  Judging from the shapes of the SEDs, most of the interlopers are
 low redshift galaxies or late-type stars. The selection of quasar candidates at $z > 3.5$ is ongoing,
 and we expect that the identification of z $\sim$ 4 quasars at these magnitude limits will be valuable
 for constraining the faint-end slope of quasars at high redshift \citep[e.g.,][]{glik10}.

\section{SUMMARY}
We have shown the characteristics of $B-,  R-$, and $I-$band data 
at Maidanak Observatory from the follow-up imaging observation of the {\it AKARI} NEP-Wide field. 
Using the SNUCAM on the 1.5m telescope, 
we covered 4.9 deg$^2$ with the 5$\sigma$ depth of $\sim$23.4,  $\sim$23.1,  
and $\sim$22.3 magnitudes (AB) at  $B-,  R-$, and $I-$band, respectively. 
We detected 63,333 sources in $B-$band, 96,460 objects in $R-$band,  
and 70,492 in $I-$band. 
These data are now being used to identifying optical counterparts for accurate astrometry 
of infrared sources and deblending of confused sources. 
Also,  our data provide a multi-wavelength coverage of the NEP-Wide field,  enabling detailed SED analysis. 
Through these SEDs,  we can get key physical properties  such as stellar mass and photometric redshifts,
 and select interesting objects such as high redshift quasars. 
Spectroscopic follow-up surveys are being conducted on infrared luminous objects 
using the information gathered from this imaging survey.

\acknowledgments
We would like to thank the observers in Maidanak Observatory 
who performed the service observation over 2 months to obtain the data. 
This work was supported by the Korea Science and Engineering Foundation (KOSEF) grant 
No. 2009-0063616,  funded by the Korea government (MEST),  
also in part by the Korea Research Foundation Grant funded by the Korean Government (MOEHRD),  
grant number KRF-2007-611-C00003.

\clearpage

\clearpage

\clearpage

\begin{deluxetable}{cccccc}
\tabletypesize{\tiny}
\tablecolumns{6}
\tablewidth{0pt}
\tablecaption{
Names of each subfield in Figure \ref{fig1} and their central positions. \label{tbl0}}
\tablehead{
\colhead{Field} & \colhead{RA} & \colhead{Dec} &\colhead{Field} & \colhead{RA} & \colhead{Dec} }
\startdata
E01 & 18:12:44.2 & 66:59:38.6 & N21 & 17:55:28.5 & 67:08:38.6 \\
E02 & 18:10:23.5 & 66:59:38.6 & N22 & 17:52:47.6 & 67:08:38.6 \\
E03 & 18:12:44.2 & 66:43:38.6 & N23 & 17:50:06.7 & 67:08:38.6 \\
E04 & 18:10:23.5 & 66:43:38.6 & S01 & 18:11:54.0 & 65:58:38.6 \\
E05 & 18:12:44.2 & 66:27:38.6 & S02 & 18:09:13.1 & 65:58:38.6 \\
E06 & 18:10:23.5 & 66:27:38.6 & S03 & 18:06:32.2 & 65:58:38.6 \\
E07 & 18:12:44.2 & 66:11:38.6 & S04 & 18:03:51.3 & 65:58:38.6 \\
E08 & 18:10:23.5 & 66:11:38.6 & S05 & 18:01:10.4 & 65:58:38.6 \\
W01 & 17:49:36.5 & 67:01:38.6 & S06 & 17:58:29.5 & 65:58:38.6 \\
W03 & 17:49:36.5 & 66:45:38.6 & S07 & 17:55:48.6 & 65:58:38.6 \\
W04 & 17:47:15.8 & 66:45:38.6 & S08 & 17:53:07.7 & 65:58:38.6 \\
W05 & 17:49:36.5 & 66:29:38.6 & S09 & 17:50:26.8 & 65:58:38.6 \\
W06 & 17:47:15.8 & 66:29:38.6 & S10 & 17:47:45.9 & 65:58:38.6 \\
W07 & 17:49:36.5 & 66:13:38.6 & S11 & 18:10:53.6 & 65:44:38.6 \\
W08 & 17:47:15.8 & 66:13:38.6 & S12 & 18:08:12.7 & 65:44:38.6 \\
N01 & 18:06:22.1 & 67:40:38.6 & S13 & 18:05:31.8 & 65:44:38.6 \\
N02 & 18:03:41.2 & 67:40:38.6 & S14 & 18:02:51.0 & 65:44:38.6 \\
N03 & 18:01:00.3 & 67:40:38.6 & S15 & 18:00:10.1 & 65:44:38.6 \\
N04 & 17:58:19.4 & 67:40:38.6 & S16 & 17:57:29.2 & 65:44:38.6 \\
N05 & 17:55:38.5 & 67:40:38.6 & S17 & 17:54:48.3 & 65:44:38.6 \\
N06 & 17:52:57.7 & 67:40:38.6 & S18 & 17:52:07.4 & 65:44:38.6 \\
N07 & 18:10:03.4 & 67:24:38.6 & S19 & 17:49:26.5 & 65:44:38.6 \\
N08 & 18:07:22.5 & 67:24:38.6 & S20 & 18:09:23.1 & 65:30:38.6 \\
N09 & 18:04:41.6 & 67:24:38.6 & S21 & 18:06:42.2 & 65:30:38.6 \\
N10 & 18:02:00.7 & 67:24:38.6 & S22 & 18:04:01.3 & 65:30:38.6 \\
N11 & 17:59:19.8 & 67:24:38.6 & S23 & 18:01:20.4 & 65:30:38.6 \\
N12 & 17:56:38.9 & 67:24:38.6 & S24 & 17:58:39.6 & 65:30:38.6 \\
N13 & 17:53:58.0 & 67:24:38.6 & S25 & 17:55:58.7 & 65:30:38.6 \\
N14 & 17:51:17.1 & 67:24:38.6 & S26 & 17:53:17.8 & 65:30:38.6 \\
N15 & 18:11:33.9 & 67:13:38.6 & S27 & 17:50:36.9 & 65:30:38.6 \\
N16 & 18:08:53.0 & 67:08:38.6 & S28 & 18:05:21.8 & 65:16:38.6 \\
N17 & 18:06:12.1 & 67:08:38.6 & S29 & 18:02:40.9 & 65:16:38.6 \\
N18 & 18:03:31.2 & 67:08:38.6 & S30 & 18:00:00.0 & 65:16:38.6 \\
N19 & 18:00:50.3 & 67:08:38.6 & S31 & 17:57:19.1 & 65:16:38.6 \\
N20 & 17:58:09.4 & 67:08:38.6 & S32 & 17:54:38.2 & 65:16:38.6 \\
\enddata
\end{deluxetable}

\clearpage

\begin{deluxetable}{cccccc}
\tabletypesize{\tiny}
\rotate
\tablecolumns{6}
\tablewidth{0pt}
\tablecaption{ Observation summary.\label{tbl1}}
\tablehead{
\colhead{Filter} & \colhead{Total exposure time [min]}  & \colhead{Exposure time per image [min]} &
\colhead{Seeing [$\arcsec$]} & \colhead{5$\sigma$ depth [AB mag]} 
& \colhead{50$\%$ Completeness [AB mag]} 
}
\startdata
$B$ & 20--45 (21) & 2--15  & 0.8--2.1 (1.4) &22.56--24.34 (23.40) & 22.05--23.86 (22.98)\\ 	
$R$&18--40 (30) & 2--10  & 0.8--1.7 (1.2) &22.07--24.14 (23.09) & 21.75--23.57 (22.60) \\ 	
$I$ & 16--64 (20) & 2 & 0.8--1.5 (1.1) &21.55--22.90 (22.33) & 21.16--22.37 (21.85) \\ 	
\enddata
\tablecomments{The values in the parentheses are the median values.}
\end{deluxetable}

\begin{deluxetable}{cccccccccccccccccc}
\tabletypesize{\tiny}
\rotate
\tablecolumns{18}
\tablewidth{0pc}
\tablecaption{ 
Photometric zero-point, seeing, 5$\sigma$ detection limit, magnitude of the completeness 50$\%$, 
	and magnitude of the reliability 99$\%$ for each subfield.\label{tbl2}}
\tablehead{
\colhead{} &  \multicolumn{5}{c}{$B-$band} &\colhead{}&  \multicolumn{5}{c}{$R-$band} &\colhead{}& 
\multicolumn{5}{c}{$I-$band} \\
\cline{2-6} \cline{8-12}  \cline{14-18} \\
\colhead{Field} & 
\colhead{$Z_p$\tablenotemark{a}}& \colhead{Seeing} & \colhead{Depth\tablenotemark{b}} & \colhead{comp50$\%$\tablenotemark{c}} & \colhead{relia99$\%$\tablenotemark{d}} & \colhead{} &
\colhead{$Z_p$\tablenotemark{a}}& \colhead{Seeing} & \colhead{Depth\tablenotemark{b}}  & \colhead{comp50$\%$\tablenotemark{c}} & \colhead{relia99$\%$\tablenotemark{d}} & \colhead{} &
\colhead{$Z_p$\tablenotemark{a}}& \colhead{Seeing} & \colhead{Depth\tablenotemark{b}} & \colhead{comp50$\%$\tablenotemark{c}} & \colhead{relia99$\%$\tablenotemark{d}}  \\
\colhead{} & \colhead{[mag]}& \colhead{[$\arcsec$]} & \colhead{[mag]} & \colhead{[mag]} & \colhead{[mag]} & \colhead{} &
\colhead{[mag]}& \colhead{[$\arcsec$]} & \colhead{[mag]} & \colhead{[mag]} & \colhead{[mag]} & \colhead{} &
\colhead{[mag]}& \colhead{[$\arcsec$]} & \colhead{[mag]} & \colhead{[mag]} & \colhead{[mag]} } 
\startdata
E01&22.88&0.84&23.95&23.63&23.62&&23.42&1.59&22.83&22.25&22.87&&23.32&1.01&22.16&21.79&21.53\\
E02&22.90&0.97&23.73&23.37&23.73&&23.39&1.63&22.95&22.29&23.17&&23.32&1.11&21.94&21.61&21.53\\
E03&22.90&0.94&23.75&23.38&23.71&&23.21&1.13&23.01&22.48&22.67&&23.32&1.11&21.98&21.69&21.12\\
E04&22.87&1.19&23.85&23.51&23.45&&23.36&1.42&22.73&22.10&22.92&&23.32&1.09&22.12&21.81&21.02\\
E05&22.85&1.49&23.18&22.80&23.54&&23.32&1.48&22.74&22.31&22.70&&23.32&0.97&22.11&21.66&20.91\\
E06&22.84&1.91&23.07&22.56&23.00&&23.34&1.49&22.79&22.30&22.66&&23.32&1.02&22.05&21.79&20.82\\
E07&22.86&1.44&23.40&22.88&23.22&&23.28&1.68&22.79&22.04&22.82&&23.32&1.15&21.84&21.16&20.78\\
E08&22.85&1.52&23.40&22.98&23.57&&23.42&1.69&23.29&22.82&23.03&&23.32&1.18&21.71&21.29&20.79\\
W01&22.88&1.97&22.97&22.41&22.89&&23.40&1.72&22.58&22.02&22.31&&23.37&0.92&22.44&21.97&21.72\\
W03&22.87&1.73&23.19&22.89&22.80&&23.39&1.60&22.65&22.28&22.45&&23.37&1.00&22.42&21.88&21.41\\
W04&22.84&1.73&23.14&22.77&22.81&&23.39&1.32&23.09&22.57&22.78&&23.37&1.02&22.33&21.84&21.55\\
W05&22.81&1.20&23.46&23.08&23.00&&23.39&1.34&23.01&22.70&22.53&&23.37&0.97&22.43&21.96&21.73\\
W06&22.78&1.42&23.16&22.81&23.01&&23.41&1.42&22.63&22.23&22.74&&23.36&1.00&22.36&21.75&21.48\\
W07&22.84&1.55&23.21&22.81&23.09&&23.41&1.33&22.98&22.55&22.13&&23.36&0.96&22.40&22.11&21.26\\
W08&22.82&1.47&23.06&22.75&22.28&&23.40&1.43&22.95&22.60&22.16&&23.36&0.95&22.38&22.11&21.54\\
N01&22.80&1.67&22.65&22.23&22.02&&23.40&1.45&23.17&22.75&22.95&&23.36&0.98&22.36&22.04&21.63\\
N02&22.76&1.71&23.00&22.48&22.79&&23.39&1.03&22.78&22.34&21.87&&23.36&1.08&22.14&21.79&21.52\\
N03&22.68&1.34&23.35&22.95&23.11&&23.39&1.09&22.40&21.92&22.01&&23.40&1.10&22.34&21.84&21.95\\
N04&22.65&0.99&23.29&23.06&22.41&&23.38&1.03&22.68&22.02&21.35&&23.40&1.17&22.29&21.83&21.79\\
N05&22.61&1.35&23.15&22.82&23.17&&23.38&0.86&22.91&22.32&21.80&&23.40&1.05&22.48&22.03&21.91\\
N06&22.57&1.32&23.23&22.91&22.58&&23.37&1.02&22.62&22.31&21.37&&23.40&1.03&22.56&22.23&21.98\\
N07&22.71&1.74&23.08&22.57&22.95&&23.36&1.04&22.51&22.19&21.46&&23.40&1.07&22.26&21.85&21.96\\
N08&22.69&1.44&23.22&22.79&22.72&&23.32&1.36&22.28&21.81&21.19&&23.40&0.98&22.34&21.82&22.08\\
N09&22.80&1.25&23.40&22.88&23.36&&23.31&1.15&22.43&21.79&21.17&&23.40&1.11&22.27&21.89&21.45\\
N10&22.78&1.69&22.98&22.51&23.24&&23.31&1.26&22.38&21.89&20.77&&23.38&1.18&22.60&22.09&22.11\\
N11&22.74&1.32&23.28&22.95&22.96&&23.30&1.30&22.32&21.92&20.86&&23.38&1.06&22.52&22.11&22.17\\
N12&22.70&1.32&23.36&23.03&22.99&&23.29&1.17&22.26&21.78&20.34&&23.38&1.13&22.31&21.87&22.22\\
N13&22.84&1.51&23.21&22.75&23.20&&23.30&1.24&22.37&21.93&21.77&&23.38&1.06&22.60&22.25&22.38\\
N14&22.80&2.13&22.83&22.30&22.71&&23.34&1.19&22.65&22.26&22.00&&23.38&1.04&22.50&22.00&21.84\\
N15&22.77&2.13&22.66&22.05&22.79&&23.34&1.26&22.43&21.75&21.77&&23.31&1.03&22.49&22.23&21.82\\
N16&22.85&1.52&23.32&22.84&23.29&&23.33&1.19&22.53&22.15&21.34&&23.31&0.96&22.33&22.02&21.26\\
N17&22.80&1.52&23.25&22.85&22.70&&23.32&1.21&22.35&21.84&21.75&&23.37&1.27&22.36&21.74&21.23\\
N18&22.81&1.49&23.38&23.05&23.15&&23.31&1.14&22.07&21.76&21.46&&23.34&1.18&22.34&21.74&21.71\\
N19&22.72&1.34&22.73&22.36&21.52&&23.34&1.29&22.88&22.31&22.45&&23.33&1.13&22.18&21.87&21.46\\
N20&22.72&1.43&23.06&22.76&22.70&&23.33&1.27&22.73&22.25&22.40&&23.37&1.13&22.33&22.10&20.99\\
N21&22.77&1.84&23.02&22.54&23.31&&23.33&1.27&22.78&22.31&22.06&&23.37&1.07&22.43&21.97&21.46\\
N22&22.77&1.71&22.56&22.17&22.43&&23.32&1.23&22.82&22.42&21.90&&23.36&1.06&22.44&22.13&20.61\\
N23&22.77&1.86&22.91&22.45&22.96&&23.31&1.27&22.63&22.19&21.79&&23.39&0.90&22.90&22.23&21.96\\
S01&22.96&1.58&23.72&23.18&23.76&&23.54&1.07&23.89&23.48&23.16&&23.42&0.88&22.84&22.37&20.77\\
S02&22.94&1.45&23.80&23.28&23.82&&23.55&1.04&23.82&23.26&23.65&&23.14&1.52&22.20&21.79&21.93\\
S03&23.10&1.23&24.10&23.43&24.18&&23.55&1.01&23.80&23.04&23.61&&23.43&0.80&22.85&21.94&19.94\\
S04&22.96&1.32&24.02&23.62&23.62&&23.55&0.93&23.90&23.28&23.78&&23.12&1.41&22.03&21.52&21.36\\
S05&23.09&1.19&24.34&23.86&23.10&&23.56&0.88&24.14&23.53&23.19&&23.42&1.23&22.40&21.78&21.39\\
S06&23.10&1.05&24.30&23.77&24.16&&23.56&0.84&24.06&23.38&23.82&&23.12&1.18&22.33&21.76&21.53\\
S07&23.10&1.44&24.00&23.56&24.12&&23.56&0.91&23.86&23.14&23.85&&23.06&1.40&22.23&21.81&21.60\\
S08&22.99&1.55&23.82&23.34&24.12&&23.56&0.92&23.92&23.57&23.86&&23.03&1.50&21.61&21.28&21.25\\
S09&23.00&1.35&23.93&23.44&24.18&&23.55&1.01&23.91&23.37&23.72&&23.14&1.06&22.37&22.09&21.97\\
S10&23.00&1.24&24.09&23.75&23.96&&23.55&1.08&23.81&23.23&23.60&&23.03&1.08&21.81&21.57&21.55\\
S11&23.00&1.39&24.08&23.71&23.78&&23.48&0.99&23.63&23.18&23.30&&22.73&1.07&21.58&21.28&20.86\\
S12&23.00&1.36&24.00&23.36&24.03&&23.48&0.92&23.80&23.27&23.70&&22.68&1.10&22.06&21.78&20.73\\
S13&23.00&1.37&24.04&23.48&23.50&&23.48&0.98&23.77&23.04&23.61&&22.78&1.09&21.89&21.35&19.30\\
S14&23.00&1.41&23.97&23.62&23.96&&23.45&1.08&23.67&23.25&23.17&&23.29&1.04&22.51&22.08&21.58\\
S15&22.98&1.31&23.81&23.37&23.97&&23.45&0.90&23.73&23.17&23.50&&22.18&1.18&21.55&21.21&20.28\\
S16&22.97&1.36&23.91&23.44&23.61&&23.45&0.84&23.82&23.37&23.50&&23.28&1.09&22.60&22.24&21.77\\
S17&22.98&1.29&23.79&23.41&23.68&&23.45&0.85&23.60&23.14&23.45&&23.27&1.10&22.51&21.86&21.00\\
S18&22.98&1.53&23.63&23.19&23.98&&23.45&0.92&23.75&23.23&23.11&&23.21&1.11&22.40&22.00&21.51\\
S19&22.97&1.59&23.57&23.11&23.44&&23.44&1.02&23.45&23.06&23.35&&23.21&1.01&22.40&21.98&21.64\\
S20&22.97&1.45&23.71&23.30&23.55&&23.38&0.95&23.60&23.01&23.24&&23.16&1.32&21.87&21.36&21.40\\
S21&22.96&1.68&23.45&22.96&23.35&&23.37&0.93&23.35&22.93&23.15&&23.24&1.39&21.83&21.32&21.51\\
S22&23.05&1.49&23.72&23.26&23.72&&23.37&0.91&23.41&22.94&22.91&&23.33&1.34&21.93&21.51&21.82\\
S23&23.04&1.14&23.89&23.34&23.87&&23.38&1.09&23.50&22.92&23.51&&23.04&1.32&21.74&21.29&21.30\\
S24&22.91&1.15&23.66&23.26&23.98&&23.38&1.13&23.55&23.22&23.18&&23.35&1.05&22.29&21.75&21.55\\
S25&22.89&1.06&23.72&23.31&23.66&&23.37&1.26&23.46&23.03&23.08&&23.36&1.07&22.12&21.77&21.63\\
S26&22.62&1.58&23.29&22.87&23.01&&23.38&1.17&23.24&22.82&22.74&&23.36&1.07&22.24&21.94&21.23\\
S27&22.83&1.00&23.63&23.23&23.36&&23.47&1.25&23.62&22.99&23.40&&23.36&1.13&22.23&21.78&21.04\\
S28&22.82&0.96&23.64&23.28&23.50&&23.33&1.46&23.21&22.59&23.19&&23.36&1.08&22.25&21.95&21.39\\
S29&22.98&1.97&22.92&22.41&23.57&&23.37&1.29&23.19&22.83&23.08&&23.36&1.11&22.22&21.92&21.26\\
S30&22.52&1.88&23.02&22.42&23.07&&23.48&1.48&23.25&22.76&23.47&&23.35&1.03&22.24&21.93&21.40\\
S31&22.94&1.47&23.18&22.82&22.81&&23.34&1.37&23.21&22.64&23.25&&23.35&1.00&22.33&21.80&21.61\\
S32&22.92&1.49&23.24&22.87&22.42&&23.42&1.32&23.11&22.65&22.85&&23.35&1.03&22.27&21.95&21.58\\

\enddata
\tablenotetext{a}{The apparent magnitude is $M=Z_p - 2.5\log$ $(DN/sec)$.}
\tablenotetext{b}{The 5$\sigma$ detection limit for a circular aperture with diameter of 3 times FWHM}
\tablenotetext{c}{The magnitude corresponding to the completeness 50$\%$ }
\tablenotetext{d}{The magnitude corresponding to the reliability 99$\%$ }
\end{deluxetable}

\begin{deluxetable}{cccccccccccccccccccc}
\tabletypesize{\tiny}
\setlength{\tabcolsep}{0.02in}
\rotate
\tablecaption{$B-,  R-$, and $I-$ band merged catalog.\label{tbl3}}
\tablewidth{0pt}
\tablehead{
\colhead{ID} & \colhead{RA} & \colhead{Dec} & \colhead{$B_{AUTO}$} & \colhead{Error} & 
\colhead{$B_{APER}$} & \colhead{Error} & \colhead{$R_{AUTO}$} & \colhead{Error} & 
\colhead{$R_{APER}$} & \colhead{Error} & \colhead{$I_{AUTO}$} & \colhead{Error} & 
\colhead{$I_{APER}$} & \colhead{Error} & \colhead{Stellarity} & \colhead{$A_B$} & 
\colhead{$A_R$} & \colhead{$A_I$} & \colhead{flags}}
\startdata
 
 E01\_00001&18:11:06.3&+66:57:11.5&22.88& 0.12&23.00& 0.10&99.00&99.00&99.00&99.00&99.00&99.00&99.00&99.00&0.89&0.18&0.11&0.08&\\
 E01\_00002&18:11:08.9&+66:51:35.9&21.67& 0.10&22.35& 0.08&99.00&99.00&99.00&99.00&99.00&99.00&99.00&99.00&0.00&0.18&0.11&0.08&\\
 E01\_00003&18:11:09.3&+67:07:40.4&23.07& 0.16&23.32& 0.14&99.00&99.00&99.00&99.00&99.00&99.00&99.00&99.00&0.03&0.19&0.12&0.08&near\_bright\_obj\\
 E01\_00004&18:11:09.3&+66:51:19.9&23.59& 0.33&24.10& 0.27&99.00&99.00&99.00&99.00&99.00&99.00&99.00&99.00&0.18&0.18&0.11&0.08&galaxy\\
 E01\_00005&18:11:09.6&+67:07:54.8&23.39& 0.25&23.96& 0.24&99.00&99.00&99.00&99.00&99.00&99.00&99.00&99.00&0.64&0.19&0.12&0.08&near\_bright\_obj\\
 E01\_00006&18:11:09.6&+66:51:22.9&23.03& 0.21&24.02& 0.25&99.00&99.00&99.00&99.00&99.00&99.00&99.00&99.00&0.69&0.18&0.11&0.08&galaxy\\
 E01\_00007&18:11:10.2&+67:07:32.0&20.85& 0.08&22.52& 0.09&99.00&99.00&99.00&99.00&99.00&99.00&99.00&99.00&0.00&0.19&0.12&0.08&near\_bright\_obj\\
 E01\_00008&18:11:10.5&+67:07:51.7&21.90& 0.11&23.19& 0.13&99.00&99.00&99.00&99.00&99.00&99.00&99.00&99.00&0.01&0.19&0.12&0.08&near\_bright\_obj\\
 E01\_00009&18:11:11.2&+67:05:29.6&22.82& 0.18&23.59& 0.18&99.00&99.00&99.00&99.00&99.00&99.00&99.00&99.00&0.32&0.19&0.12&0.08&\\
 E01\_00010&18:11:11.2&+66:50:56.2&99.00&99.00&99.00&99.00&22.15& 0.12&22.21& 0.12&99.00&99.00&99.00&99.00&0.36&0.18&0.11&0.08&near\_edge\\
 E01\_00011&18:11:11.3&+66:50:53.4&99.00&99.00&99.00&99.00&21.91& 0.11&22.04& 0.12&99.00&99.00&99.00&99.00&0.47&0.18&0.11&0.08&near\_edge\\
 E01\_00012&18:11:11.4&+67:05:27.2&22.35& 0.14&23.33& 0.15&99.00&99.00&99.00&99.00&99.00&99.00&99.00&99.00&0.21&0.19&0.12&0.08&galaxy\\
 E01\_00013&18:11:11.4&+67:07:10.7&20.55& 0.06&20.60& 0.06&99.00&99.00&99.00&99.00&99.00&99.00&99.00&99.00&0.98&0.19&0.12&0.08&near\_bright\_obj\\
 E01\_00014&18:11:11.4&+66:50:51.2&99.00&99.00&99.00&99.00&22.77& 0.16&22.27& 0.14&99.00&99.00&99.00&99.00&0.46&0.18&0.11&0.08&near\_edge\\
 E01\_00015&18:11:11.5&+66:51:19.2&99.00&99.00&99.00&99.00&22.69& 0.17&22.46& 0.16&99.00&99.00&99.00&99.00&0.43&0.18&0.11&0.08&near\_edge\\
 E01\_00016&18:11:11.6&+67:05:27.1&22.95& 0.17&23.66& 0.19&99.00&99.00&99.00&99.00&99.00&99.00&99.00&99.00&0.71&0.19&0.12&0.08&\\
 E01\_00017&18:11:11.7&+67:07:12.8&20.50& 0.06&20.62& 0.06&99.00&99.00&99.00&99.00&99.00&99.00&99.00&99.00&0.96&0.19&0.12&0.08&near\_bright\_obj\\
 E01\_00018&18:11:11.7&+66:51:04.3&99.00&99.00&99.00&99.00&22.81& 0.19&22.53& 0.18&99.00&99.00&99.00&99.00&0.40&0.18&0.11&0.08&near\_edge\\
 E01\_00019&18:11:11.9&+67:07:22.9&23.59& 0.79&23.24& 0.14&99.00&99.00&99.00&99.00&99.00&99.00&99.00&99.00&0.00&0.19&0.12&0.08&near\_bright\_obj\\
 E01\_00020&18:11:12.1&+66:53:23.3&99.00&99.00&99.00&99.00&18.10& 0.05&19.19& 0.05&99.00&99.00&99.00&99.00&0.03&0.18&0.11&0.08&\\

\enddata
\tablecomments{
(1) The identification number of the source. 
Each ID consists of the subfield name (Figure \ref{fig1}) and number in the order of their RA.\\
(2) and (3) The J2000.0 right ascension (RA) and the declination (Dec) of a source in a sexigesimal.\\
(4)--(15) are the total magnitudes, its magnitude errors, aperture magnitudes and its magnitude errors of a source 
in the $B-, R-$, and $I-$band. The aperture magnitudes are obtained with the aperture diameter of 3 times FWHM.
The magnitudes are not corrected for the extinctions.\\ 
(16) The stellarity index of a source from SExtractor.  See Figure \ref{fig6} for the distribution of this index. \\
(17--19) The galactic extinction values of $B-, R-,$ and $I-$band from extinction map of \citet{schl98}. \\
(20) The flags parameter. `near$\_$bright$\_$obj' for the spurious detection near bright stars, 
`near$\_$edge' for the spurious detection near the edge of a single image, 
or  `galaxy' for the non-stellar sources classified at section \ref{sec:galcut}.
}
\end{deluxetable}

\begin{deluxetable}{ccccccccccccccc}
\tabletypesize{\tiny}
\rotate
\tablecolumns{15}
\tablewidth{0pc}
\tablecaption{ 
Galaxy and stellar number counts. \label{tbl4}}
\tablehead{
\colhead{} &  \multicolumn{4}{c}{$B-$band} &\colhead{}&  \multicolumn{4}{c}{$R-$band} &\colhead{}& 
\multicolumn{4}{c}{$I-$band} \\
\cline{2-5} \cline{7-10}  \cline{12-15} \\
\colhead{mag} & \colhead{galaxy}& \colhead{} & \colhead{corrected} &\colhead{star} & \colhead{} &
\colhead{galaxy}& \colhead{} & \colhead{corrected} &\colhead{star}& \colhead{} &
\colhead{galaxy}& \colhead{} & \colhead{corrected} &\colhead{star}\\
\colhead{} & \colhead{number}& \colhead{error} & \colhead{number} & \colhead{number}& \colhead{} &
\colhead{number}& \colhead{error} & \colhead{number} & \colhead{number}& \colhead{} &
\colhead{number}& \colhead{error} & \colhead{number}& \colhead{number}} 
\startdata
15.75&3.68E+00&9.85E-01&3.68E+00&1.46E+02&&1.13E+01&1.73E+00&1.13E+01&2.24E+02&&9.21E+00&1.56E+00&9.21E+00&2.72E+02\\
16.25&3.42E+00&9.49E-01&3.42E+00&1.74E+02&&1.71E+01&2.12E+00&1.71E+01&2.96E+02&&1.76E+01&2.15E+00&1.76E+01&3.35E+02\\
16.75&6.58E+00&1.32E+00&6.58E+00&2.19E+02&&2.58E+01&2.61E+00&2.58E+01&3.27E+02&&3.30E+01&4.28E+00&3.31E+01&4.03E+02\\
17.25&1.26E+01&1.82E+00&1.26E+01&2.49E+02&&4.24E+01&3.34E+00&4.24E+01&3.91E+02&&5.93E+01&5.91E+00&5.95E+01&5.04E+02\\
17.75&2.13E+01&2.37E+00&2.13E+01&2.92E+02&&8.58E+01&4.75E+00&8.58E+01&4.58E+02&&1.03E+02&9.62E+00&1.03E+02&5.84E+02\\
18.25&2.97E+01&2.80E+00&2.97E+01&3.42E+02&&1.15E+02&8.29E+00&1.15E+02&5.25E+02&&1.95E+02&1.42E+01&1.97E+02&7.06E+02\\
18.75&5.55E+01&3.82E+00&5.55E+01&3.73E+02&&2.19E+02&1.27E+01&2.20E+02&5.99E+02&&3.34E+02&3.08E+01&3.38E+02&7.96E+02\\
19.25&8.05E+01&4.60E+00&8.05E+01&3.91E+02&&3.73E+02&2.81E+01&3.78E+02&7.24E+02&&5.07E+02&7.27E+01&5.30E+02&8.81E+02\\
19.75&1.27E+02&5.77E+00&1.27E+02&4.78E+02&&6.00E+02&5.98E+01&6.20E+02&8.28E+02&&7.29E+02&1.02E+02&7.87E+02&9.12E+02\\
20.25&1.99E+02&2.62E+01&2.02E+02&5.38E+02&&9.25E+02&9.37E+01&9.78E+02&9.36E+02&&1.08E+03&1.24E+02&1.24E+03&9.83E+02\\
20.75&3.76E+02&5.74E+01&3.89E+02&6.09E+02&&1.26E+03&1.20E+02&1.38E+03&1.02E+03&&1.54E+03&1.49E+02&2.01E+03&1.05E+03\\
21.25&6.59E+02&8.96E+01&7.10E+02&7.69E+02&&1.48E+03&1.38E+02&1.73E+03&1.00E+03&&2.02E+03&1.70E+02&3.12E+03&9.80E+02\\
21.75&9.65E+02&1.12E+02&1.11E+03&8.38E+02&&1.71E+03&1.47E+02&2.21E+03&8.55E+02&&\nodata&\nodata&\nodata&\nodata\\             
22.25&1.41E+03&1.35E+02&1.88E+03&9.24E+02&&2.36E+03&1.71E+02&3.37E+03&7.24E+02&&\nodata&\nodata&\nodata&\nodata\\             
22.75&2.01E+03&1.58E+02&2.97E+03&9.31E+02&&3.37E+03&2.02E+02&5.56E+03&6.31E+02&&\nodata&\nodata&\nodata&\nodata\\             
23.25&3.13E+03&1.95E+02&5.09E+03&7.14E+02&&6.12E+03&2.73E+02&1.12E+04&3.98E+02&&\nodata&\nodata&\nodata&\nodata\\             

\enddata
\tablecomments{
Units of the galaxy and stellar number densities and their errors are Numbers 0.5mag$^{-1}$ deg$^{-2}$. 
}
\end{deluxetable}

\end{document}